\documentclass[a4paper,UKenglish,cleveref, autoref, thm-restate]{lipics-v2021}
\newcommand{\op}[1]{\texttt{#1}}
\usepackage{amsmath}
\usepackage[inline]{enumitem}
\usepackage{soul}
\usepackage{placeins}

\hideLIPIcs

\long\def\full#1{%
{#1}%
}
\long\def\arxiv#1{%
{#1}%
}
\long\def\elsevier#1{%
{}%
}
\long\def\conf#1{%
{}%
}

\title{Space-Efficient Graph Coarsening with Applications to Succinct Planar
Encodings%
} 

\author{Nina Hammer}{THM, University of Applied Sciences Mittelhessen, Giessen,
Germany }{nina.hammer@mni.thm.de}{https://orcid.org/0009-0009-3600-2087}{}{}{}
\author{Frank Kammer}{THM, University of Applied Sciences Mittelhessen, Giessen,
Germany }{frank.kammer@mni.thm.de}{https://orcid.org/0000-0002-2662-3471}{}{}{}
\author{Johannes Meintrup}{THM, University of Applied Sciences Mittelhessen,
Giessen, Germany
}{johannes.meintrup@mni.thm.de}{https://orcid.org/0000-0003-4001-1153}{Funded by
the Deutsche Forschungsgemeinschaft (DFG, German Research Foundation) --
379157101. }{}{}

\authorrunning{N. Hammer and F. Kammer and J. Meintrup}

\Copyright{N. Hammer and Frank Kammer and Johannes Meintrup}



\category{} 

\relatedversion{} 

\EventEditors{\ } \EventNoEds{1}
\EventLongTitle{\ }
\EventShortTitle{\ } \EventAcronym{\ } \EventYear{2022}
\EventDate{May 2022} \EventLocation{\ } 
\EventLogo{} \SeriesVolume{1} \ArticleNo{1}
\begin{document}

\nolinenumbers
\maketitle

\begin{abstract}
    We present a novel space-efficient graph coarsening technique for $n$-vertex
    planar graphs $G$, called \textit{cloud partition}, which partitions the
    vertices $V(G)$ into disjoint sets $C$ of size $O(\log n)$ such that each $C$
    induces a connected subgraph of $G$. Using this partition $\mathcal{P}$ we
    construct a so-called \textit{structure-maintaining minor} $F$ of $G$ via
    specific contractions within the disjoint sets such that $F$ has $O(n/\log n)$
    vertices. The combination of $(F, \mathcal{P})$ is referred to as a
    \textit{cloud decomposition}.
    
    For planar graphs we show that a cloud decomposition can be constructed in
    $O(n)$ time and using $O(n)$ bits. Given a cloud decomposition $(F,
    \mathcal{P})$ constructed for a planar graph $G$ we are able to find a balanced
    separator of $G$ in $O(n/\log n)$ time. Contrary to related publications, we do
    not make use of an embedding of the planar input graph. We generalize our cloud
    decomposition from planar graphs to $H$-minor-free graphs for any fixed graph
    $H$. This allows us to construct the succinct encoding scheme for $H$-minor-free
    graphs due to Blelloch and Farzan (CPM 2010) in $O(n)$ time and $O(n)$ bits
    improving both runtime and space by a factor of $\Theta(\log n)$.

    As an additional application of our cloud decomposition we show that, for
    $H$-minor-free graphs, a tree decomposition of width  $O(n^{1/2 + \epsilon})$
    for any $\epsilon > 0$ can be constructed in $O(n)$ bits and a time linear in
    the size of the tree decomposition. A similar result by Izumi and Otachi (ICALP
    2020) constructs a tree decomposition of width $O(k \sqrt{n} \log n)$ for graphs
    of treewidth $k \leq \sqrt{n}$ in sublinear space and polynomial time.
    
    \full{Finally, we implemented our cloud decomposition algorithm and
        experimentally verified its practical effectiveness on both randomly
        generated graphs and real-world graphs such as road networks. The
        obtained data shows that a simplified version of our algorithms
        suffices in a practical setting, as many of the theoretical worst-case
        scenarios are not present in the graphs we encountered.}
\end{abstract}

\section{Introduction}
\label{sec:introduction}

Graphs are used to model a multitude of systems that can be expressed via
entities and relationships between these entities. Many real-world problems
operate on very large graphs for which standard algorithms and data structures
use too much space. This has spawned an area of research with the aim of
reducing the required space. Examples include large
road-networks~\cite{strasser_et_al:LIPIcs:2020:12947} or social graphs arising
from interactions between users of large internet
communities~\cite{Floreskul_Tretyakov_Dumas_2014}. Therefore, it is of high
interest to find compact representations of such graphs, space-efficient
algorithms and other space-efficient or succinct data structures. In the
following we denote by $n$ the number of vertices of a graph under consideration
{and $m$ the number of edges.} An algorithm is called
space-efficient if it has (almost) the same asymptotic runtime as a standard
algorithm for the same problem, but uses asymptotically fewer bits. Examples
include space-efficient graph searching algorithms such as depth-first search
and breadth-first search, which run in linear time, but use $O(n \log n)$ bits
with standard methods. Space-efficient solutions lower the space requirement to
$O(n)$ bits and keep the runtime asymptotically (almost) the same. For a data
structure or algorithm to be called {\it succinct} the space used must be $Z +
o(Z)$ bits with $Z$ being the information theoretic minimum to store the data.
One of the most researched topics regarding space-efficiency are graph-traversal
algorithms such as depth-first search (DFS). It is still an open research
question if a linear-time DFS exists that uses $O(n)$ bits, with the current
best bound of $O(n + m + \textrm{min}\{n, m\}\log^*n)$ due to
Hagerup~\cite{hagerup2020}, with $\log^*n$ being the iterated logarithm. This
result is the work of gradual improvements over the span of seven years spanning
multiple publications by multiple research groups, with some of the first
results providing a runtime $O(m \log n)$ time due to Asano et
al.~\cite{asano2014} and a runtime of $O((n+m) \log \log n)$ due to Elmasry et
al.~\cite{elmasry2015}. Other typical
problems that can be easily solved in settings where space is of no concern,
such as storing spanning trees or simple mappings between vertices, require new
problem-specific strategies when considered in the space-efficient setting.

{Another interesting setting is aiming for arbitrary polynomial time
runtime, but using only $o(n)$, or even $O(\log n)$ bits. In such settings
information can be recomputed in polynomial time, and for problems such as
mappings this is often trivial, while other problems such as deciding if two
vertices are connected by a path in a graph requires the very involved algorithm
due to Reingold~\cite{reingold}, but is quite easy in a space-efficient
setting.} Thus, 
our 
space-efficient setting 
differentiates itself quite strongly from both, the
well-researched sublinear-space settings and the common settings that do not
regard space as limited.

Many large graphs that arise in practice have some known structural property
that allows the design of specialized techniques that make use of these
properties. A common such structural property is the existence of small
separators. Such a {\it separator} is a small subset of vertices whose removal
disconnects the graph if it was connected previously, or increases the number of
connected components if it was not connected. {For the graphs of interest to
this work the separators are so-called balanced separators. For now the
intuition suffices that such balanced separators split the graph in somewhat
equally parts. Section~\ref{sec:background} contains precise definitions of
these terms and also defines what we consider small in regard to separators.
Arguably the most well-known graphs that contain such small balanced separators
are planar graphs, which are graphs that can be drawn in the plane without
overlapping edges. All planar graphs contain a balanced separator of size
$O(\sqrt n)$~\cite{doi:10.1137/0136016}.} Similar separator theorems exist for
almost-planar graphs~\cite{10.1137/0216064} (which includes road networks) and
well-formed meshes such as nearest neighbor graphs~\cite{10.1145/256292.256294}
and $H$-minor-free graphs for some fixed graph $H$~\cite{5670816}.
For arbitrary separable graphs there exists a polylogarithmic approximation
algorithm for finding balanced separators due to Leighton and Rao~\cite{21958}.

We present a novel partitioning scheme called \textit{cloud partition} for
planar graphs that partitions the vertices of a {connected} input graph $G$ into
connected subsets called \textit{clouds} that induce a connected subgraph and
are of size $O(\log n)$. For a cloud partition $\mathcal{P}$ constructed for $G$
we construct a so-called \textit{structure-maintaining minor} $F$ of $G$ induced
by $\mathcal{P}$. For easier reading comprehension we call vertices of such a
minor \textit{nodes}. Intuitively, a node $v\in V(F)$ is mapped to one or more
clouds $C \in \mathcal{P}$ such that $|V(F)|=O(n/\log n)$---the exact mapping
is outlined in Section~\ref{sec:graphcoursening}. A solution for some problems
such as finding small separators can be found in $F$ and then translated to an
approximate solution for $G$. As $F$ contains $O(n/\log n)$ nodes, the time
and space bounds of linear or superlinear algorithms can be decreased by a
factor of $\Omega(\log n)$ when being executed on $F$ instead of $G$. We show
how this speedup is especially helpful for recursive algorithms such as the
recursive separator search used during the computation of the succinct
representation of separable graphs due to Blelloch and
Farzan~\cite{10.5555/1875737.1875750}. Additionally, we show that small
modifications of this recursive separator search can be used to find a tree
decomposition of width $O(n^{1/2+\epsilon})$ for any $\epsilon > 0$ for planar
graphs in $O(n)$ bits and a time linear in the size of the tree decomposition.
Finally, we generalize our partitioning scheme from {planar graphs
to $H$-minor-free graphs for any fixed graph $H$}.

{One of the key points of our novel scheme is that we do not make use of a
(planar) embedding of the input graph, as there is no known way to construct
such an embedding with $O(n)$ bits and in linear time. It is often implied that
an embedding is given alongside a planar graph in many of the publications
regarding planar graphs, as it is computable in linear
time~\cite{10.1145/321850.321852} when space is of no concern. Additionally, it
is often required that the planar graph is maximal. Both of these properties
are, for example, required for the major result of the $O(\sqrt{n})$-separator
theorem~\cite{doi:10.1137/0136016} or the so-called $r$-partition of Klein et
al.~\cite{10.1145/2488608.2488672}, which similarly to our result partitions the
input graph into regions of size $O(n/r)$. Again, if the graph is not maximal,
it can be easily made maximal in linear time with the help of an embedding. In a
space-efficient context we can not make use of either of these properties and
are thus quite limited. For sublinear space settings there exist algorithms that
produce an embedding in polynomial time with a rather large polynomial
degree~\cite{Allender00thecomplexity, 10.1145/2591796.2591865}. Thus, the
$O(n)$-bit setting provides a unique challenge due to the additional goal of
matching the runtime of non space-efficient algorithms.}

Succinct and space-efficient representations of planar graphs is a highly
researched topic partially due to the practical applications. For compressing
planar graphs without regards to providing fast access operations refer to the
work of Keeler et al.~\cite{10.1016/0166-218X(93)E0150-W} for an $O(n)$ bits
representation and for a compression within the information theoretic lower
bound refer to He et al.~\cite{DBLP:journals/corr/cs-DS-0101021}. Due to Munro
and Raman~\cite{doi:10.1137/S0097539799364092} there exists an encoding using
$O(n)$ bits that allows constant-time queries which has subsequently been
improved by Chiang et al.~\cite{10.5555/365411.365518} to use a constant factor
less space. We use the succinct representation due to Blelloch and Farzan, which
allows encoding arbitrary separable graphs and subsequently allows constant-time
adjacency-, neighborhood- and degree-queries~\cite{10.5555/1875737.1875750},
which builds on the work of Blanford et al.~\cite{10.5555/644108.644219}.
{For $H$-minor-free graphs for a fixed graph $H$, their encoding
takes $\Theta(n \log n)$ time and $\Theta(n \log n)$ bits using their described
technique of recursive separator searches. We are able to improve it to $O(n)$
time and $O(n)$ bits. For planar graphs we only improve the space-requirement
from $\Theta(n \log n)$ to $O(n)$ due to the algorithm of
Goodrich~\cite{10.1006/jcss.1995.1076}\full{ which can be used instead of the generic
recursive separator search described by Blelloch and Farzan}}. Note that all
mentioned publications above regarding succinct representations of planar graphs
use $\Theta(n \log n)$ bits during the construction. Additionally, they assume a
planar embedding is given explicitly or implicitly by making direct use of it,
or referring to other publications as sub-routines that require them. For
maximal planar graphs there exist special encodings. Some major research in this
field is due to Aleardi et al.~\cite{DBLP:journals/jocg/AleardiD18,
DBLP:conf/cccg/AleardiDS05, 10.1007/11534273_13,
DBLP:journals/tcs/AleardiDS08}\conf{.}\full{, which deals with practical and
theoretical representations of such maximal planar graphs, typically arising in
applications making use of geometric meshes or computer graphics. Note that
these space-efficient encodings of maximal planar graphs also make use of and
store an embedding, as navigation of the faces is a vital feature in these
applications.}

The general technique of graph coarsening has been heavily used in the past for
practical and theoretical algorithms~\cite{chevalier09, pmlr-v119-fahrbach20a,
10.1145/2459976.2459984, 1383165, 10.1007/978-3-319-38851-9_20}. A typical
approach is to find a matching $E' \subset E(G)$ for a graph $G$ and contract
the edges in $E'$ to find a minor $F$ of $G$. The problem at hand is then solved
on $F$ and the solution is translated to an approximate solution for
$G$~\cite{1383165}. Highly specialized practical algorithms using sophisticated
data structures are implemented in the METIS~\cite{Karypis98afast} and
SCOTCH~\cite{10.1007/3-540-61142-8_588} library. Other approaches such as graph
coarsening on the GPU~\cite{auer2012graph, auer2012gpu} or graph coarsening via
neural networks~\cite{cai2021graph} have been developed. \full{Note that the use
of sophisticated data structures that focus on runtime optimization do not lend
themselves to modifications to make them space-efficient. Thus, there was a need
to develop a novel partitioning scheme for this work, in particular a scheme
that does not require an embedding when dealing with planar graphs.} 

For finding tree decompositions in a space-efficient manner refer to the work of
Kammer et al.~\cite{DBLP:journals/corr/abs-1907-00676} and the work of Izumi and
Otachi~\cite{izumi_et_al:LIPIcs:2020:12474}. {Izumi and Otachi showed that for a
given graph $G$ with treewidth $k \leq \sqrt{n}$ there exists an algorithm that
obtains a tree decomposition of width $O(k \sqrt{n}\log{n})$ in polynomial time
and $O(k \sqrt{n} \log^2 n)$ bits. Note that the polynomial degree in the
runtime is rather large due to the use of the well known $s$-$t$ reachability
algorithm of Reingold~\cite{reingold}. Izumi and Otachi additionally mention
that for planar graphs there exists a polynomial-time and sublinear-space
algorithm for finding a tree decomposition of a planar graph due to a simple
recursive separator search with the result of Imai et al.~\cite{6597770}, which
present an algorithm for finding balanced separators of size $O(\sqrt{n})$ in
planar graphs with sublinear space and polynomial runtime. Note that the
algorithm of Imai et al.~makes use of an embedding of the input graph and also
uses the algorithm of Reingold.} In contrast, we present an algorithm that
computes a tree decomposition of width $O(n^{1/2 +\epsilon})$ in time linear in
the size of the tree decomposition using $O(n)$ bits. As the tree decomposition
has $O(n)$ bags of size $O(n^{1/2 +\epsilon})$ this results in a runtime of
$O(n^{3/2 +\epsilon})$.

In Section~\ref{sec:background} we present concepts needed to understand the
subsequent sections. In Section~\ref{sec:graphcoursening} we present our
space-efficient graph-coarsening framework. Following that, in
Section~\ref{sec:applications} and~\ref{sec:applications2}  
we show how this scheme is used for finding separators, tree decompositions and
constructing succinct encodings of planar graphs. In Section~\ref{sec:general}
we generalize our work to $H$-minor-free graphs. \full{
Finally, in Section~\ref{sec:pract} we analyze the practical effectiveness of our
decomposition algorithm. We implemented the cloud partition algorithm 
and tested it on both randomly generated graphs, and real-world
data such as street networks. 
}
\conf{For space reasons most
proofs can be found in the appendix.}
\section{Background}
\label{sec:background}

We operate in the standard word RAM model of computation with word size $w =
\Omega(\log n)$. This assumes the existence of read-only input memory,
read/write working memory and write-only output memory. When we talk about
space-usage we focus on bits used in the read/write working memory.
\full{Sometimes this model is also referred to as the \textit{register input
model}~\cite{DBLP:journals/jcss/Frederickson87}. }{This is a common
setting for space-efficient algorithms.}

We make use of common graph theoretic notations and terminology. Refer to
Diestel~\cite{Diestel07graphtheory} for more information. When using
space-efficient graph algorithms the exact representation of the input graph is
important. We assume that any input graph is given via \textit{adjacency arrays}
or an equivalent interface. Given a vertex $u$ and index $i \leq \op{degree}(u)$
this allows us to determine in constant time the $i$th edge $\{u, v\}$ out of
$u$. All our input graphs are assumed to be undirected with the vertices labeled
from $1,\ldots, n$. Such a labeling is given implicitly by the order of the
adjacency arrays. Given the use of adjacency arrays we store each undirected
edge $\{u, v\}$ as two directed \textit{arcs} $uv$ and $vu$. For each arc $uv$
we call $vu$ the \textit{co-arc of $uv$}. We assume the existence of so-called
\textit{crosspointers} that allow us to find the co-arc of an arc $uv$ in
constant time. Typically, this is realized by storing an index $i$ in addition
to the vertex $v$ in $u$'s adjacency array. The index $i$ then indicates the
position of $u$ in $v$'s adjacency array.
We assume w.l.o.g. that every given graph $G$ is connected
as otherwise all our techniques can be done iteratively for each connected
component of $G$. {We assume that a given input graph is in
read-only memory.}

A \textit{separator $S$} of a graph $G=(V, E)$ is a subset of $V$ such that its
removal from $V$ divides $V$ into non-empty sets $A \subset V$ and $B \subset V$
so that $\{A, S, B\}$ is a partition of $V$ with the constraint that all paths
from a vertex $u \in A$ to a vertex $v \in B$ contain at least one vertex of
$S$. If  $|A| < \alpha n$ and $|B| < \alpha n$ for some
$\alpha<1$, then $S$ is called a {\it{$\alpha$-balanced separator}} or simply 
{\it{balanced separator}}.

{A family of graphs $\mathcal{G}$ that is closed under taking
vertex-induced subgraphs satisfies the {\it{$f(\cdot)$-separator theorem}}
exactly if, for constants $\alpha < 1$ and $\beta > 0$, each member $G \in
\mathcal{G}$ has an $\alpha$-separator $S$ of size $|S| < \beta f(n)$. We say a
family of graphs $\mathcal{G}$ is \textit{separable} exactly if it satisfies the
{\it{$n^c$-separator theorem}} for some $c<1$. We say a graph is separable if it
belongs to a separable family of graphs. For planar graphs there exist an
$O(\sqrt{n})$-separator theorem with $\alpha=2/3$ that runs in linear
time~\cite{doi:10.1137/0136016}, later extended to graphs of bound genus with
the same runtime~\cite{GILBERT1984391}. For minor-closed graph classes excluding
the complete graph $K_t$ on $t$ vertices for some constant $t$ there exists an
$O(n^{(2-\epsilon)/3}$)-separator theorem with runtime
$O(n^{1+\epsilon}+m)$~\cite{5670816}. Note that this includes $H$-minor-free
graphs for any fixed graph $H$.

The definition of the separable graph families are not closed under taking
minors, but vertex-induced subgraphs~\cite{10.5555/1875737.1875750}. As we
specifically construct minors $F$ of a separable graph $G$ at various points in
this publication, we require the more restrictive property that $F$ also belongs
to the same family of separable graphs as $G$, which holds for all graph classes
explicitly mentioned in the previous paragraph. }

Due to Lipton et al.~\cite{Lipton1977GeneralizedND} is it known that a separable
graph class $\mathcal{G}$ has \textit{bound density} $d$ for some constant $d$.
This means that each $G\in\mathcal{G}$ contains at most $dn$ edges. In
particular, we use the well-known fact, by Euler's Formula, that for a planar
graph $G=(V, E)$ it holds that $|E| \leq 3 |V| - 6$. For planar bipartite graphs
a stronger bound $|E| \leq 2|V|-4$ holds.

A \textit{tree decomposition} of a graph $G=(V, E)$ consists of a
tree $T$ and a family $\mathcal{X}$ of subsets $X_w$ (called \textit{bags}) of
$V$, one for each $w \in V(T)$, such that:
\begin{enumerate}
    \item $\bigcup_{w \in V(T)} X_w=V$
    \item for all $\{u, v\} \in E$ there exists $w \in V(T)$ such that $u, v \in
    X_w$
    \item for all $w_1, w_2, w_3 \in V(T)$, if there is a path from $w_1$ to
    $w_3$ that
    contains $w_2$, then $X_{w_1} \cap X_{w_3} \subseteq X_{w_2}$.
\end{enumerate}

The \textit{width}
of a tree decomposition is the size of the largest bag minus one. The treewidth
of a graph $G$ is the smallest width amongst all possible tree decompositions of
$G$. 


We make use of \textit{indexable dictionaries}, which is a structure that
supports constant-time rank-select queries over a bitvector. The $\op{rank}(i)$
operation counts the number of occurrences of bits set to $1$ before the $i$th
index and the $\op{select}(i)$ operation returns the index of the $ith$ bit set
to $1$. 

\begin{lemma}[\cite{10.1145/1290672.1290680}]
    Given a bitvector $S$ of length $\ell$ there is an indexable dictionary on
    $S$ that requires $o(\ell)$ additional bits, supports rank-select queries in
    constant time and can be constructed in $O(\ell)$ time.
\end{lemma}

What can be thought of a dynamic alternative to indexable dictionaries is the
so-called \textit{choice dictionary}~\cite{DBLP:conf/mfcs/Hagerup19}. A choice
dictionary is initialized for a universe $1,\ldots, \ell$ and supports constant
time $\op{insert}$, $\op{delete}$, and $\op{contains}$ operations, {
with the latter returning $\op{true}$ exactly if an element of the universe is
contained in the choice dictionary.} Additionally, the choice dictionary offers
the operation $\op{choice}$ that returns an arbitrary member and iteration over
its members. The iteration outputs all members of a choice
dictionary in constant time per member. All operations run in constant time and
the iteration over its members is linear in the number of members. The following
lemma has been adapted from~\cite{kammer_et_al}, with some minor rephrasing to
make it more clear in the context of this paper.

\begin{lemma}
    There is a succinct choice dictionary initialized for the universe $1,\ldots, \ell$
    that occupies $\ell+o(\ell)$ bits and provides constant-time
    $\op{insert}$, $\op{delete}$, $\op{contains}$ and $\op{choice}$ operations
    and constant-time (per member) iteration. The choice dictionary can be
    initialized in $O(\ell)$ time. 
\end{lemma}

We make use of a folklore technique called \textit{static space allocation}
allowing us to store $\ell$ items of varying size compactly. The following
description is adapted from~\cite{DBLP:journals/corr/KammerKL16}. Each item
$B_k$ occupies $d_k$ bits for $k \in \{1,\ldots, \ell\}$. Denote by $L$ the
amount of bits all these items totally occupy. We want to store all these items with $L
+ o(L)$ bits such that we can access each item $B_k$ in $O(1)$ time for $k \in
\{1,\ldots, \ell\}$. In $O(\ell + L)$ time compute the sums $s_k = k +
\sum_{j=1}^{k-1}d_j$ and an indexable dictionary over a bitvector $S$ of size
$\ell + L$ with $S[i]=1$ exactly if $i=s_k$ for $k \in \{1,\ldots, \ell\}$. The
location of $B_k$ is then equivalent to $\op{select}_B(k)-k$. Refer
to~\cite{10.1007/978-3-319-42634-1_10, HAGERUP201916,
DBLP:journals/corr/KammerKL16} for more information and detailed
descriptions. 

For traversing graphs we use standard breadth-first search (BFS), which puts a
start vertex in a first queue, then the unprocessed neighbors of the first queue
in a second queue, swaps the queues and repeats. It can easily be seen that
space requirement is $O(n\log n)$ bits. More precise, the BFS uses $O(\ell \log
n)$ bits where $\ell$ is the maximum number of vertices in a queue at any point
of the BFS. 

In this work we need to store subgraphs $G'$ of a given graph $G=(V, E)$.
Similar to the techniques used by Hagerup et al.~\cite{HAGERUP201916} we are
able to store $G'$ using $O(n+m)$ bits. We store a subgraph $G'$ of $G$ via $n$
choice dictionaries $D_v$ for each vertex $v \in V(G)$. Each $D_v$ has length
$d_v$ with $d_v$ being the degree of $v$ in $G$. The choice dictionaries are
stored using static space allocation. A member $i$ in $D_v$ then indicates the
existence of the $i$th arc out of $v$'s adjacency array. Another choice
dictionary $D'$ of length $n$ can be used to mark vertices of degree $>0$.
Iterating over the neighborhood of a vertex $v \in V(G')$ can be done in linear
time of the degree of $v$ in $G'$. This additionally allows dynamic insertion
and deletion of edges as long as $G'$ remains a subgraph of $G$, i.e., no new
edges can be added. Note that we do not directly allow the deletion of vertices.
Instead, when a vertex has degree $0$ in $G'$ (due to the deletion of all
incident edges) it is marked in $D'$. This allows to iterate over all vertices
$V'=\{v \in V(G') | d_v > 0\}$ in $O(|V'|)$ time with $d_v$ being the degree of
$v$ in $G'$. Additionally, an arbitrary vertex $v \in V(G')$ with $d_v > 0$ can
be obtained via $D'.\op{choice}()$. We refer to such a structure as a
\textit{dynamic subgraph $G'$ of $G$}. Note that a dynamic subgraph $G'$ can be
used to direct an edge $\{u, v\} \in E(G')$ by deleting only the arc $uv$ or
$vu$ in $G'$. Using this we are able to implement the next lemma in $O(n)$ time
and $O(n)$ bits using the same algorithm as described
in~\cite{10.5555/644108.644219}. 
\conf{The proof can be found in Appendix~\ref{sec:proofbound}.}
\newcommand{\lemboundindegree}{
\begin{lemma}
    \label{lem:boundindegree}
    Let $G$ be a separable graph. We can obtain the directed graph $G'$ of $G$
    such that each vertex of $G'$ has bounded in-degree (out-degree) in $O(n)$
    time and $O(n)$ bits.
\end{lemma}
}
\lemboundindegree
\newcommand{\proofboundindegree}{
\begin{proof}
We implement a space-efficient algorithm of the algorithm outlined by Blandford
et al.~\cite{10.5555/644108.644219}. Note that for all separable graph classes
there exist some constant $b$ such that at least half of the vertices have
degree $\leq2b$~\cite{10.5555/644108.644219} and each separable graph $G=(V, E)$
has $E=O(|V|)$~\cite{Lipton1977GeneralizedND}. The algorithm due to Blanford et
al. takes as input a separable graph $G=(V, E)$, obtains the vertices $V'
\subset V$ such that each $v \in V'$ has degree $\leq 2b$. All edges of $E$ with
an endpoint in $V'$ are then directed towards $V'$. Then all vertices of $V'$
are removed, and the process is repeated until all edges are handled this way.
The runtime is $O(n)$ as $>1/2|V|$ vertices are removed in each step. We now
describe the technical details of how to obtain such a directed graph in $O(n)$
time and $O(n)$ bits.

Let $G$ be a separable graph and $G_h, G'$ dynamic subgraphs of $G$, both
initially containing all edges of $G$. Note that each undirected edge $\{u,v\}
\in E$ is represented by two arcs $uv$ and $vu$. Removing one of these arcs can
then be thought of as directing this {undirected edge from $u$ to
$v$ or $v$ to $u$}. In the following we construct a directed version of $G$
iteratively in $G'$. We use $G_h$ as a helping structure during the
construction. As an additional structure we use a choice dictionary $Q$ which
functions as a queue for vertices of degree $\leq 2b$. Iterate over all vertices
$v$ of degree $>0$ in $G_h$ using the choice dictionary $D'$
(Section~\ref{sec:background}). For each $v$ obtain the degree $d_v$ of $v$ by
iterating over the adjacency list of $v$ in $G_h$. If $d_v \leq 2b$, then add
$v$ to $Q$. Finding and adding all these vertices to $Q$ takes $O(n+m)=O(n)$
time. In $G'$ remove all arcs $uv$ with $u \in Q$ and $v \neq Q$. In the case
that both $u$ and $v$ are in $Q$ remove the arc $uv$ with $u < v$. Removing all
these arcs in $O(n)$ time. For each $v \in Q$ remove all incident edges from
$G_h$ and continue the iteration in $G_h$ until all edges are handled. The
runtime is the same as due to Blanford et al. while the use of dynamic subgraphs
uses $O(n)$ bits.
\end{proof}
}
\full{\proofboundindegree}

\section{Graph Coarsening Framework for Planar Graphs}
\label{sec:graphcoursening}
In this section we outline a strategy for coarsening a planar graph $G=(V, E)$.
The idea is to create a specific type of partition $\mathcal{P}$ of $V$, which
we call \textit{cloud partition} that induces a unique minor $F$ of $G$, defined
later. We refer to each $C \in \mathcal{P}$ as a \textit{cloud}. 
Thus, a cloud is a set of vertices.
In the
following we describe the exact specifications that define such a cloud
partition. For each $C \in \mathcal{P}$, $C$ induces a connected subgraph of $G$
and $|C| \leq \lceil c \log n \rceil$ for an arbitrary, but fixed constant $c$. We
differentiate between different types of clouds $C \in \mathcal{P}$, which we define after
introducing some terminology. Let $C_1, C_2 \in \mathcal{P}$ with $C_1 \neq
C_2$. We call an edge $\{u, v\} \in E$ a \textit{border edge} exactly if $u \in
C_1$ and $v \in C_2$. We then refer to $C_1$ and $C_2$ as \textit{adjacent} or
\textit{neighbors} and as \textit{incident to $\{u, v\}$}. Furthermore, we call
$C$ a \textit{big cloud} if $|C| = \lceil c \log n \rceil$ and a \textit{small
cloud} otherwise. We call a small cloud $C$ a \textit{leaf cloud} if $C$ is
adjacent to one cloud, a \textit{bridge cloud} if it is adjacent to two clouds
and a \textit{critical cloud} if it is adjacent to at least three clouds. We
call two clouds $C_1, C_2$ adjacent to a bridge cloud $B$ \textit{connected
by $B$}. A cloud partition is created by starting with the
following scheme: Initially, mark all vertices as unvisited. Run a BFS from an
arbitrary unvisited vertex $v$. The BFS only traverses vertices marked unvisited.
Each time an unvisited vertex is traversed, it is marked visited. The BFS runs
until either $\lceil c \log n \rceil$ vertices are marked visited or no
unvisited vertices can be reached by the BFS. In the first case a big cloud is
found and in the second case a small cloud is found. Repeat this until no
unvisited vertices remain, always starting a new BFS at an arbitrary unvisited
vertex. Only a partition created in this way is referred to as a cloud
partition. 
By fixing the search for unvisited vertices and the BFS algorithm, each graph has a fixed cloud
partition.
The next two observations are derived directly from the process of
creating a cloud partition.

\begin{observation}
    \label{lem:nosmalladjacent}
    Let $\mathcal{P}$ be an arbitrary cloud partition created for a graph $G$.
    Then no two small clouds $C_1, C_2 \in \mathcal{P}$ are adjacent.
\end{observation}

\begin{observation}
    \label{obs:numbigcloudsbound}
    Let $\mathcal{P}$ be an arbitrary cloud partition created for a graph $G$.
    Since each big cloud has size $\lceil c \log n \rceil$ for a constant
    $c$, $\mathcal{P}$ contains at most $n / \lceil c \log n \rceil$ big
    clouds.
\end{observation}

For planar graphs, the number of critical clouds can be bounded similarly. The
proof is based on the fact that a planar graph has $O(n)$ edges.

\newcommand{\lemcritical}{
\begin{lemma}
    \label{lem:critical}
    Let $\mathcal{P}$ be a cloud partition created for a planar graph $G$
    containing $k$ big clouds. Then $\mathcal{P}$ contains $O(k)$ critical
    clouds.
\end{lemma}
}
\lemcritical
\newcommand{\proofcritical}{
\begin{proof}
    Consider the graph $F=(V', E')$ constructed by contracting the vertices of
    each cloud $C \in \mathcal{P}$ to a single vertex $v \in C$. The graph $F$
    is a minor of $G$ with $|V'| = |\mathcal{P}|$ and each vertex $v \in V'$
    represents a single cloud $C_v \in \mathcal{P}$. Two vertices $v, u \in V'$
    are adjacent in $F$ exactly if $C_v$ and $C_u$ are adjacent. For brevity's
    sake we refer to vertices of $F$ introduced for critical clouds as critical
    vertices, and vertices introduced for big clouds as big vertices. Assume
    that all vertices that are not big vertices or critical vertices are removed
    in $F$ and all edges between big vertices are removed as well. This turns
    $F$ into a bipartite planar graph due to
    Observation~\ref{lem:nosmalladjacent}, { with big vertices on
    one side of the bipartition, and small and critical on the other}. 
    It is well-known that $|E'|\leq 2|V'|-4$ for bipartite planar graphs. 
    Denote by $\ell$ the number of critical vertices and $k$
    the number of big vertices in $F$. Note that for each critical vertex $v \in
    V'$ it holds $\op{degree}(v) \geq 3$. As such, $|E'| \geq 3\ell$. Using this
    the number of critical vertices in $F$ is bounded via $3\ell \leq 2|V'|-4 =
    2k+2\ell-4$ and thus $\ell \leq 2k -4$. (Note that the addition of any edges
    between big vertices in $F$ can only tighten this bound.) As the number of
    critical vertices in $F$ is equal to the number of critical clouds in
    $\mathcal{P}$ we have shown that there are at most $2k - 4$ critical clouds
    in $\mathcal{P}$.
\end{proof}
}
\proofcritical

The next corollary follows together with Obs.~\ref{obs:numbigcloudsbound}.

\begin{corollary}
    \label{cor:boundcrit}
    Let $\mathcal{P}$ be a cloud partition created for a planar graph $G$. Then
    $\mathcal{P}$ contains $O(n / \log n)$ critical clouds.
\end{corollary}

Note that for leaf or bridge clouds no such bound exists as a cloud partition
can contain $O(n)$ leaf and bridge clouds. Extreme examples include a cloud
partition $\mathcal{P}$ with one big cloud $C$ and $n - |C|$ leaf clouds, all
adjacent to $C$ and containing only one vertex. For bridge clouds a similar
example partition $\mathcal{P}$ can be constructed, with two big clouds $C_1,
C_2 \in \mathcal{P}$ and $n - (|C_1|+|C_2|)$ bridge clouds, all adjacent to both
$C_1$ and $C_2$.

Next we focus on the construction of a specific weighted minor $F$ of $G$ with
weights $w(v)$ assigned to each node $v \in V(F)$. {Intuitively,
$F$ represents a minor of $G$ that is constructed by repeatedly contracting the
vertices in one or more clouds, and the weights keep track of the number of
vertices that have been contracted. We call such an $F$ a
\textit{structure-maintaining minor} of $G$, with the formal definition outlined
shortly.} 
 As each such minor is constructed specifically for a cloud partition $P$ we say
that $F$ is \textit{induced by $P$}. We now define the properties of such a
minor and follow with a sketch of a construction. Let $F$ be a
structure-maintaining minor induced by a cloud partition $\mathcal{P}$. Denote
by $\mathcal{C}_u$ the set of clouds contracted to a node $u \in V(F)$. Each
node $u \in V(F)$ is assigned a weight $w(u)=\sum_{C \in \mathcal{C}_u}|C|$. For
each $u$ one of the following two properties holds: (1) $|\mathcal{C}_u|=1$,
$G[C]$ is connected for $C \in \mathcal{C}_u$ and $w(u)=\lceil c \log n \rceil$
for some fixed constant $c$ or (2) each $C\in \mathcal{C}_u$ is adjacent to
exactly the same clouds, $G[\bigcup_{C \in \mathcal{C}_u}C]$ contains
$|\mathcal{C}_u|$ connected components. Additionally, $\{v, w\} \in E(F)$
exactly if there exist adjacent clouds $C_v$ and $C_w$ with $C_v \in
\mathcal{C}_v$ and $C_w \in \mathcal{C}_w$. 

We now outline the construction of $F$. Initially, $F = (V'=\emptyset,
E'=\emptyset)$. {For each cloud $C$ that is big or critical, add a
node $v$ to $V'$ with $w(v)=|C|$}. Add edges between $u, v \in V'$ to $E'$
exactly if the clouds $u$ and $v$ represent are adjacent. We call $v$ a
\textit{big node} if it was added to $V'$ for a big cloud and \textit{critical
node} if it was added for a critical cloud. For each pair of big clouds $C_1,
C_2 \in \mathcal{P}$ connected via one or more bridge clouds, let
$\mathcal{B}_{C_1, C_2} \subset \mathcal{P}$ be the set of all such bridge
clouds. Add a single node $v$ to $V'$ adjacent to the nodes $u, w$ added to $V'$
for $C_1, C_2$ and set $w(v)=\sum_{B \in \mathcal{B}_{C_1, C_2}}|B|$. We refer
to such a node $v$ as a \textit{meta-bridge node}. For each big cloud $C \in
\mathcal{P}$ adjacent to one or more leaf clouds, denote by $\mathcal{L}_{C}
\subset \mathcal{P}$ the set of all leaf clouds adjacent to $C$. For each such
$C$ add a single node $v$ to $V'$ adjacent to the node $u$ added to $V'$ for $C$
with $w(v) = \sum_{L \in \mathcal{L}_{C}}|L|$. We call such a node $v$ a
\textit{meta-leaf node}. 

Since we have only $O(n/\log n)$ big nodes, we also have only $O(n/\log n)$
critical, meta-bridge and meta-leaf nodes, {expressed explicitly in
the following lemma}. Note that the weight of a meta-bridge or meta-leaf node
can be bounded only by $n$.
We next bound the number of nodes and edges 
of $F$.

\newcommand{\lemsizeoff}{
\begin{lemma}
    \label{lem:sizeoff}
    Let $\mathcal{P}$ be a cloud partition constructed for a planar graph $G$
    and $F=(V', E')$ the structure-maintaining minor induced by $\mathcal{P}$. Then
    $|V'| = O(n/\log n)$ and $O(|E'|)=O(|V'|)$.
\end{lemma}
}
\lemsizeoff
\newcommand{\proofsizeoff}{
\begin{proof}
{As $F$ is a minor of $G$ it is planar as well and so it holds that
$O(|E'|)=O(|V|')$. From Obs.~\ref{obs:numbigcloudsbound} we know that the number
of big nodes in $F$ is $k=O(n/ \log n)$. From the construction of meta-leaf
nodes it directly follows that each meta-leaf node is adjacent to exactly one
big node and vice-versa. Therefore there are at most $k$ meta-leaf nodes in
$V'$. For meta-bridge nodes the proof follows analogous, consider two big nodes
$u, v$ that are adjacent to the same meta-bridge node $w$. Then there is no
meta-bridge node $w\neq w$ adjacent to both $u$ and $v$, as per the
construction. By the fact that $F$ is planar it follows that there are at most
$O(k)=O(n/ \log n)$ meta-bridge nodes in $V'$. Putting it all together we arrive
at $|V'|=O(n/ \log n)$. }
\end{proof}
}
\proofsizeoff

In the following we describe a data structure for a cloud partition and show how
it can be constructed and stored in $O(n)$ time and $O(n)$ bits. We refer to
this data structure as \textit{cp-structure}. A cp-structure constructed for a
planar connected graph $G$ admits the following operations:
\begin{itemize}
    \item $\op{type}(v)$: Given a vertex $v$ outputs the type of cloud $v$
    belongs to (big, small, critical, bridge or leaf) in $O(1)$ time.
    \item $\op{cloud}(v)$: Given a vertex $v$ returns all vertices of the cloud
    $C$ in which $v$ is contained in $O(|C|)$ time and $O(|C|\log n)$ space.
    \item $\op{border}(v, k)$: Outputs if the $k$-th arc out of $v$'s adjacency
    array is part of a border edge in $O(1)$ time, with $v \in V(G)$.
\end{itemize}

Additionally, the structure allows access to the subgraph $G'$ of $G$ induced by
all non-border edges of $\mathcal{P}$. The graph $G'$ admits adjacency array
access to its vertices and edges. { The following lemma describes
the runtime and space usage of constructing a cloud partition and a
cp-structure. \conf{The proof can be found at
Appendix~\ref{sec:proofcpstruct}.}}
%

\newcommand{\lemcpstruct}{
\begin{lemma}
    \label{lem:cpstruct}
    Let $G$ be a planar connected graph. We can compute a cloud partition
    $\mathcal{P}$ of $G$ and a cp-structure of $\mathcal{P}$ in $O(n)$ time and
    $O(n)$ bits.
\end{lemma}
}
\lemcpstruct
\newcommand{\proofcpstruct}{
\begin{proof}
    In the following we construct a variety of data structures that are combined
    to a final cp-structure. { For each type of cloud, construct a
    bitvector of the same name of length $n$, i.e., \op{big}, \op{small},
    \op{critical}, \op{bridge} or \op{leaf}. During the following procedure a
    bit at index $i$ will be set to $1$ in the respective bitvector to indicate
    that the vertex $i \in V(G)$ is of the respective type, with all bits set to
    $0$ initially.} To track border edges of $\mathcal{P}$ we maintain a dynamic
    subgraph $G'$ of $G$, initially containing all edges of $G$ as outlined in
    Section~\ref{sec:background}. A final bitvector $\op{start}$ of length $n$
    is used during the construction, containing a $1$ at index $i$ exactly if
    $i$ was the first vertex discovered in a cloud during the initial
    construction of the cloud partition $\mathcal{P}$.

    The first step of the construction is computing $\mathcal{P}$ by finding all
    clouds. Starting at an arbitrary unvisited vertex $v$ set $\op{start}[v]=1$
    and run a BFS until $\lceil c \log n \rceil$ unvisited vertices are visited,
    or no unvisited vertices can be reached. Denote by $A$ the set of all
    vertices found during such a BFS. If $|A|=\lceil c \log n\rceil$ we have
    found a big cloud. In this case, iterate over the neighborhood of each
    vertex $u \in A$ and mark all edges that are incident to a vertex $w \notin
    A$ as border edge by removing it in $G'$. Additionally, for all vertices $u
    \in A$ set $\op{big}[u]=1$. In the case that $|A| < \lceil c \log n\rceil$
    we have found a small cloud for which we set $\op{small}[u]=1$. In both
    cases we start the same procedure again from an unvisited vertex $v$. If no
    such vertices remains, we have successfully found $\mathcal{P}$ and have
    constructed the bitvectors that allows us to differentiate between small and
    big clouds. Additionally, the graph $G'$ is exactly the subgraph of $G$
    induced by non-border edges.

    The next step is to construct the bitvectors for critical, bridge and
    leaf clouds. For each vertex $v$ that is a start vertex of a big cloud $C$,
    identifiable as $\op{big}[v]=1$ and $\op{start}[v]=1$, run a BFS in $G$.
    During this BFS explore all small clouds $S$ adjacent to $C$. If the
    vertices $w \in S$ are marked as part of a critical cloud during previous
    iterations of this step, identifiable via $\op{critical}[w]=1$, ignore $S$
    completely. I.e., at no point traverse edges that are already identified as
    leading to a critical cloud. Otherwise, for each $w \in S$ set
    $\op{leaf}[w]=1$ during the first exploration of $S$, $\op{leaf}[w]=0$ and
    $\op{bridge}[w]=1$ during the second exploration and $\op{bridge}[w]=0$ and
    $\op{critical}[w]=1$ during the third. By this each small cloud and each
    border edge is accessed at most a constant number of time. Note that each
    big cloud is traversed once. Afterwards the exact type of cloud a vertex
    belongs to is accessible via the respective bitvectors in constant time.

    The operation $\op{type}(v)$ is supported in constant time by checking if
    the bit at index $v$ is set to $1$ in the respective bitvectors constructed
    for the cloud types in $O(1)$ time. The operation $\op{cloud}(v)$ can be
    accessed by running a BFS from $v$ in $G'$ until no vertex can be reached.
    As $G'$ contains no border edges it contains each cloud as a connected
    component. The runtime is linear in the number of vertices of the cloud $C$
    the vertex $v$ is contained in. The space requirement in terms of bits of a
    BFS run in a cloud $C$ is $O(|C|\log n)=O((\log n)^2)$ as this equals the
    maximum size of the queue of the BFS. The operation $\op{border}(v, k)$ can
    be accessed by simply checking if the edge is present or not in $G'$.
\end{proof}
}
\full{\proofcpstruct}

Next we show how to construct a structure-maintaining minor $F$ induced by a
cloud partition $\mathcal{P}$ of a graph $G$ in $O(n)$ time and $O(n)$ bits of
space. By Lemma~\ref{lem:sizeoff}, $F$ can be stored in $O(n)$ bits by adjacency
lists. We additionally store a bi-directional mapping from each big/critical
cloud $C$ to the node $v \in V(F)$ added to $V(F)$ to represent $C$. This
mapping is stored by a pointer at $v$ to a single vertex $v' \in C$ and a
pointer at $v'$ to $v$. As there are $O(n /\log n)$ pointers to store for this
bi-directional mapping we use standard pointers of size $\Theta(\log n)$. To
store the pointers from the direction of a vertex $v \in V(G)$ we use
static-space allocation. Note that the choice of the vertex to store these
pointers (per cloud) is arbitrary, but should be fixed. We choose the vertex
with the lowest label. For meta-bridge and meta-leaf nodes $v \in V(F)$ we store a
pointer from $v$ to the lowest labeled vertex $v'$ amongst all clouds
represented by the meta-bridge or meta-leaf node. The details are described in the
proof of the next lemma\conf{, which can be found at
Appendix~\ref{sec:proofconstructf}}. Next, we define a special operation on $F$.
\begin{itemize}
    \item $\op{expand}(v)$: Given a node $v \in V(F)$ with weight $w(v)$,
   first determine the cloud $\mathcal{C}$ mapped to $v$ and then 
   return iteratively 
   all vertices part of the clouds $C \in \mathcal{C}$.
\end{itemize}
{ Later we make use of the $\op{expand}$ operation when translating solutions
for problems such as finding separators from $F$ to a solution to $G$. The key
difficulty is implementing the $\op{expand}$ operation for a meta bridge $v$. As
the clouds represented by $v$ may induce up to $n-O(\log n)$ connected
components in $G$ and we store only a pointer to the lowest labeled vertex $u
\in V(G)$ amongst all such clouds represented by $v$, we are unable to simply
output each cloud without additional information or we must traverse (in the
worst case) the entire graph $G$. The idea is to store a spanning tree that
spans all vertices in clouds represented by $v$. As mentioned, since these
clouds each induces a connected component, a spanning tree with these clouds
does not exist. Therefore, we create a spanning tree by additionally using the
vertices of one adjacent big cloud. Traversing all vertices represented by $v$
can now be done by traversing the respective spanning tree. The key observation
for this is that we can direct the edges of $F$ in such a way that each vertex
has up to $c=O(1)$ outgoing edges (Lemma~\ref{lem:boundindegree}). By this we
are able to store all spanning trees in $O(n)$ bits due to the fact each big
cloud is used only by $c$ spanning trees.Concerning the problem of having $c$
spanning trees that use the vertices of one big cloud, i.e., that overlap, our
approach is to assign each spanning tree a color from a set of $c$ colors. These
spanning trees are then stored in $c$ dynamic subgraphs, with each subgraph
storing all spanning trees of one color. } \newcommand{\lemconstructf}{
\begin{lemma}
    \label{lem:constructf}
    Let $G$ be a connected planar graph and $\mathcal{P}$ a cloud partition of
    $G$ given as a cp-structure. We can construct the structure-maintaining
    minor $F$ induced by $\mathcal{P}$ in $O(n)$ time and bits 
such that the $\op{expand}$ operation on $v\in V(F)$
runs in $O(w(v) + \log n)$ time.
\end{lemma}
}
\lemconstructf
\newcommand{\proofconstructf}{
\begin{proof}
    We construct $F$ iteratively. Initially, we add all big and critical nodes
    $v$ and the bi-directional mapping from $v$ to one vertex $v'$ in the cloud
    $v$ represents. During this process, we maintain a bitvector to mark
    vertices of $G$ as visited. An additional bitvector $X$ is maintained for
    marking the lowest numbered vertex in each big or critical cloud. Starting
    at the lowest labeled vertex $v$, obtain the cloud $C$ with $v \in C$ via
    the $C=\mathcal{P}.\op{cloud}(v)$ operation. Mark all vertices of $C$ as
    visited, add a node $v'$ to $V(F)$ and store a pointer at $v'$ to $v$ and
    store a pointer from $v$ to $v'$ in a standard linked list $L$ and set
    $X[v]=1$.  Simply iterate this process until all big and critical nodes are
    added to $V(F)$. Note that by this construction the lowest labeled vertex of
    a cloud is always the vertex for which the bi-directional mapping is
    constructed, as we iterate over the bitvector for visited vertices in order.
    Constructing an indexable dictionary for $X$ and transforming $L$ into an
    array (in $O(|L|)$ time and $O(|L|\log n)$ bits) gives us exactly the static
    space allocation for pointers from vertices in $V(G)$ to nodes in
    $V(F)$.

    Now we add all edges between big and critical nodes. We maintain two
    bitvectors $\op{complete}$ and $\op{discovered}$ for vertices, initially
    containing only $0$ bits. In $\op{complete}$ we mark processed big clouds
    and in $\op{discovered}$ we mark clouds discovered from a big cloud during
    the iteration to ensure that these clouds are only visited once.
    Iteratively, start at an arbitrary vertex $v$ that is not marked complete
    and is contained in a big or critical cloud. Obtain the cloud $C$ via
    $\mathcal{P}.\op{cloud}(v)$. Mark all vertices of $C$ as complete. For each
    border edge $\{u, w\}$ with $u \in C$ and $w$ in a big or critical cloud
    $C'$ adjacent to $C$ and for which $w$ is not marked as discovered obtain
    $C'$ via $\mathcal{P}.\op{cloud}(w)$. Mark all vertices in $C'$ as
    discovered and obtain the node $w' \in V(F)$ that represents $C'$ via the
    bi-directional mapping, stored for the lowest labeled vertex of $C'$. Add
    the arc $v'w'$ to the adjacency list of $v'$. Note that the arc $w'v'$ is
    added during a different iteration, i.e., when we start at $C'$ and find $C$
    as an adjacent cloud. Once this process has been completed for all border
    edges incident to $C$ mark all adjacent big or critical clouds as no longer
    discovered by an analogous process. The entire process is iterated until no
    vertices $v$ in critical or big clouds remain that are not marked complete.
    The weights for each node $v$ added to $V(F)$ during this step can be
    determined and stored during the run by simply counting the vertices in the
    cloud that $v$ represents. We so have added $O(n / \log n)$ nodes and edges
    to $F$ during this process. For each edge $\{u, v\} \in E(F)$ (as the two
    arcs $uv$ and $vu$) we have explored the clouds of each endpoint a constant
    number of times via the $\op{cloud}$ operation, which runs in linear time.
    As each such cloud has size $O(\log n)$ and $|E(F)|=O(n/\log n)$ the overall
    runtime is $O(n)$.

    In the next step we add all meta-leaf nodes to $F$. As in the previous step we
    maintain the same bitvectors and re-initialize them. Iteratively, start at
    an arbitrary vertex $v$ that is not marked complete and is contained in a
    big cloud $C$, obtainable via $\mathcal{P}.\op{cloud}(v)$. For each border
    edge $\{u, w\}$ with $u \in C$ that is incident to a leaf cloud and for
    which $w$ is not marked as discovered obtain the cloud $C'$ adjacent to $C$
    via $\mathcal{P}.\op{cloud}(w)$ and mark all vertices of $C$ as discovered.
    Once all border edges have been iterated this way, add a node $u$ to $V(F)$
    incident to $v'$. The weight of $u$ is set to the number of vertices in all
    leaf clouds adjacent to $C$. Additionally, we store a pointer from $u$ to
    the lowest labeled vertex $u'$ amongst all leaf clouds which are represented
    by $u$. Repeat the entire process until no (unvisited vertices in) big
    clouds remain.
    As in the previous step, for each edge $\{u, v\}$ added to $E(F)$ we iterate
    over the clouds of $u$ and $v$ a constant number of time and therefore have
    a runtime of $O(n)$.
    
    Finally, we need to add all meta-bridge nodes to $F$. Similar as to the
    previous steps we start at a big cloud $C$ and explore all neighboring
    bridge clouds $B$ and the big clouds connected to $C$ via the bridge clouds.
    Assume the same bitvectors as in the previous steps are available. We work
    on two dynamic subgraphs $G_a, G_b$ of $G$, with $G_a$ initially being a
    copy of $G$ and $G_b$ being the empty graph. In $G_a$ we remove border edges
    incident to vertices in bridge clouds for which we have already added a
    meta-bridge node to $F$ to avoid unnecessary explorations. During the
    process we start at an arbitrary big cloud $C$ not marked complete. We then
    construct $G_b$ as the graph induced by $C$, all bridge clouds $B$ adjacent
    to $C$ and all big clouds connected to $C$ via $B$. In $G_b$ we start at an
    arbitrary bridge cloud $B$ adjacent to $C$, remove the edges incident to
    vertices of $B$ and vertices of $C$, and traverse to the big cloud $C'$
    connected to $C$ via $B$. From $C'$ we then explore all adjacent bridge
    clouds $B$, removing all traversed edges and all edges adjacent to vertices
    of $B$ in $G_b$ and $G_a$. Once finished, we add a meta-bridge node $v$ to
    $F$ and the edges $\{v, w\}$ and $\{v, u\}$ with $w$ being the node added to
    $V(F)$ for $C$ and $u$ being the node added to $V(F)$ for $C'$. The weight
    of $v$ is simply the number of vertices traversed in the bridge clouds
    represented by $v$. Additionally, we store a pointer from $v$ to the lowest
    labeled vertex $v'$ amongst all bridge clouds which are represented by $v$.
    Finally, we mark all vertices of $C$ as complete and repeat the process from
    the next big cloud not yet marked complete. For each meta-bridge node $v$
    added to $V(F)$ we explore the two clouds $C, C'$ represented by the two big
    nodes adjacent to $v$ once. Additionally, we explore all vertices in bridge
    clouds represented by $v$ a constant number of time. As we add $O(n /\log
    n)$ meta-bridge nodes overall the runtime is $O(n)$. 
    
    It remains to show how to implement the $\op{expand}$ operation. For big and
    critical nodes $v \in V(F)$ the $\op{expand}$ operation is trivially solved
    by obtaining the vertex $v'$ mapped to $v$ and calling
    $\mathcal{P}.cloud(v')$. We therefore now focus only on meta-bridge and
    meta-leaf nodes. We first give a non-technical overview of our technique
    followed by a technical implementation and analysis. Denote by
    $\mathcal{C}_u$ the set of clouds represented by a meta-leaf or meta-bridge
    node $u \in V(F)$.  
%
%
    Intuitively, our goal is to create a tree $T_u$ spanning over all
    vertices in $\cup_{C \in \mathcal{C}_u} C$.
    As the clouds in $\mathcal{C}_u$ induce disconnected components in $G$, we must
    add vertices of one cloud $C_0$ to make $T_u$ a tree where 
   $C_0\in \mathcal{P}$ is  
    one of the clouds  adjacent to all
    $C \in \mathcal{C}_u$ in $F$.
     Once such a spanning tree is available for all meta-bridge and meta-leaf
    nodes $u$, the $\op{expand}$ operation can simply be achieved by traversing
    $T_u$ via a depth-first search. It is easy to see that the height of each
    $T_u$ is $O(\log n)$ since $F[C_0 \cup \mathcal{C}_u]$ is a star so that we
    can easily execute a standard DFS and only need $O(n)$ bits.

    For meta-leaf nodes storing these spanning trees can easily be
    achieved in a single dynamic subgraph $G_{\op{leaf}}$ as all $T_u$ are
    pairwise disjoint. 

    {For meta-bridge nodes we make use of the fact that any separable
    graph can be directed in such a way that each vertex has at most in-degree
    $d$ for some graph class dependent constant $d$ as per
    Lemma~\ref{lem:boundindegree}. We first create a minor $F'$ of $F$ that
    contains only big nodes and an edge between two big nodes $u, v$ exactly if there
    is a meta-bridge node adjacent to both $u$ and $v$.
    Finally, we split each edge by adding the corresponding meta-bridge node in
    the middle of the edge and direct the two split edges in the direction of
    the original edge. }

    {Recall that each $T_u$ constructed for the bridge clouds
    $\mathcal{C}_u$ represented by some meta-bridge node $u$ spans
    only one big cloud $C_v$. We choose $C_v$ as the cloud such that there is a
    directed edge $uv$ in $F'$ between the nodes $u$ and $v$ corresponding to
    $\mathcal{C}_u$ and $C_v$, respectively.}
    By this there are at most $d$ meta-bridge nodes 
    that use the same big cloud. Observe
    that all spanning trees that do not use the same big cloud are disjoint.
    From this, we are able to store all spanning trees in $d$ dynamic subgraphs
    $G_1, \ldots, G_d$. In intuitive terms, we can color all spanning trees with
    $d$ colors such that any two spanning trees $T, T'$ are disjoint if they are
    assigned the same color. The spanning trees can be constructed
    in intermediate computations  while repeating the same process for adding
    the meta-bridge nodes $v$ to $V(F)$ and their incident edges. Analogously
    (and simpler) the same is true for meta-leaf nodes. For each meta-bridge
    node $v$ we store a pointer to the one dynamic subgraph $G_i \in G_1,
    \ldots, G_d$ in which $T_u$ is contained. To traverse $T_u$ we then obtain
    the vertex $v' \in V(G)$ mapped to $v$ and start a depth-first search from
    $v$ in $G_i$. For each $T_u$ it holds $|V(T_u)|=w(u) + O(\log n)$.
    Constructing each $T_u$ can be done in $O(w(u)+\log n)$ time, and there are
    $O(n/\log n)$ spanning trees overall, {as each node in $F$
    results in up to $d=O(1)$ spanning trees and there are $O(n/\log n)$ nodes
    in $F$. As such, it takes $O(n)$ time and uses $O(n)$ bits.}
\end{proof}
}
\full{\proofconstructf}

We refer to the combined data structure of Lemma~\ref{lem:cpstruct} and
Lemma~\ref{lem:constructf} as a \textit{cloud decomposition $(F, \mathcal{P})$
of $G$}. 

\section{Applications: Succinctly
Encoded Planar Graphs}
\label{sec:applications}
In this section we show how a cloud decomposition $(F, \mathcal{P})$ constructed
for a planar graph $G$ can be used to find a balanced separator of size
$O(\sqrt{n \log n})$ in $O(n/\log n)$ time and $O(n)$ bits, which
in turn can be used to construct a succinct encoding of planar graphs, described
later. 
Afterwards we show how such a cloud decomposition can be used together with the
search for balanced separators to compute a tree decomposition with width
$O(n^{1/2+\epsilon})$ for any $\epsilon > 0$ of a planar graph $G$. We make use
of an $O(n)$-time $O(n \log n)$-bits algorithm for finding 2/3-balanced
separators of size $O(\sqrt{n})$ in weighted planar graphs as long as no vertex
has a weight more than 1/3 times the total
weight~\cite{kleinMozes2021,doi:10.1137/0136016,Miller86}. The next theorem
shows how a balanced separator can be constructed for a planar graph with the
help of a cloud decomposition. The idea is to run a slightly modified standard
algorithm for finding balanced separators $S$ on the structure-maintaining minor
$F$ and translating $S$ to a separator $S'$ of $G$.

\newcommand{\lemseparator}{
\begin{theorem}
    \label{thm:separator}
    Let $G=(V, E)$ be a planar graph and $(F=(V', E'), \mathcal{P})$ a cloud
    decomposition of $G$. We can construct a balanced separator $S$ of $G$ with
    size $O(\sqrt{n \log n})$ in $O(n/\log n)$ time using $O(n)$ bits. 
\end{theorem}
}
\lemseparator
\newcommand{\proofseparator}{
\begin{proof}
    Start 
    to search for a separator $S\subset V'$ on the weighted graph
    $F$ in time $O(|V'|)=O(n/\log n)$ and $O(|V'|\log n)=O(n)$ bits. Denote by
    $\mathcal{S}\subset \mathcal{P}$ the clouds of $\mathcal{P}$ represented by
    nodes of $S$ and by $w(S)$ the sum of the weights $w(v)$ of all $v \in S$.
    We can construct a separator $S'$ for $G$ via $S'=\bigcup_{C\in
    \mathcal{S}}C$. We store $S'$ in a bitvector of length $n$, initialized to
    all $0$-bits in $O(n/\log n)$ time. For each $v \in S$ we obtain an iterator
    over the clouds $v$ represents via $F.\op{expand}(v)$ and simply add the
    vertices returned by the iterator to $S'$ in $O(w(s))$ time.
    Note that $|S'|=O(|S|\log n)=O(\sqrt{n/\log n}\log n)=O(\sqrt{n \log n})$.

    A balanced separator $S$ of $F$ can only be found if no node $v \in V'$ has
    weight 
    larger than $1/3$ times the total weight. Only meta-leaf or meta-bridge
    nodes can have such large weights. If such a node exists, we split it into
    up to three nodes by distributing its clouds as equally as possible to the
    three nodes and run the search on the modified graph.
%
%
\end{proof}
}
\full{\proofseparator}

Due to Blelloch and Farzan~\cite{10.5555/1875737.1875750} there exists a
succinct encoding of an arbitrary separable graph $G=(V, E)$ that provides
constant-time adjacency, degree and neighborhood queries. The encoding is
constructed via recursive application of a separator algorithm such that $V$ is
split into two sets $A$ and $B$ via a separator $S$. This recursion is continued
for $G_a=G[A \cup S]$ and $G_b=G[B \cup S]$ until the remaining graphs are of
size at most $\log^{\delta}n$ for a graph class specific constant $\delta$,
which they refer to as \textit{mini graphs}. In several papers, 
what is constructed here is referred to as a \textit{separator hierarchy}. For
technical reasons the edges between vertices of $S$ are only included in $G_a$
and not $G_b$. The graph $G$ is then encoded by a combination of these mini
graphs, which in turn are further decomposed by the same recursive separator
search into \textit{micro graphs}, which are then small enough to be handled via
lookup tables encoding every possible micro graph and in turn are used to encode
the mini graphs. Constructing these mini and micro graphs can be done in $O(n)$
time for a planar graph $G$ using the techniques described by Blelloch and
Farzan combined with the algorithm of Goodrich for constructing a separator
hierarchy~\cite{10.1006/jcss.1995.1076}, but we want to use only $O(n)$ bits and
maintain the runtime. The main result of~\cite{10.5555/1875737.1875750} is
summed up by the next theorem. 

\begin{theorem}[\cite{10.5555/1875737.1875750}]
    \label{thm:encoding}
    Any family of separable graphs with entropy $\mathcal{H}(n)$ can be
    succinctly encoded in $\mathcal{H}(n)+o(n)$ bits such that adjacency,
    neighborhood, and degree queries are supported in constant time.
\end{theorem}

One important point of the construction is the fact that the number of
\textit{duplicate vertices} in each mini graph is bounded. A duplicate is a
vertex that is contained in more than one mini graph. A vertex becomes a
duplicate any time it is part of a separator $S$, as it is then contained in
both $G_a$ and $G_b$. The exact bound is defined by the next lemma.

\begin{lemma}[\cite{10.5555/1875737.1875750}]
    \label{lem:blelloch}
    The number of mini graphs is $\Theta(n/\log^{\delta}n)$. The total number
    of duplicates among mini graphs is $O(n/\log^2n)$. The sum of number of
    vertices of mini graph together is $n + O(n/\log^2n)$.
\end{lemma}

Our goal is to execute the recursive decomposition of the graph $G$ via a cloud
decomposition $(F, \mathcal{P})$ by recursively searching for separators in $F$
until the entire weight of the remaining graphs is less than $\log^{\delta}n$.
These small subgraphs of $F$ then are expanded to be exactly the mini graphs of
$G$ we aim to find. For each mini graph a new cloud decomposition is
constructed, and the recursive separator search is repeated for each mini graph
to construct the micro graphs. \conf{The proof can be found in
Appendix~\ref{sec:proofminig}.}

\newcommand{\lemminig}{
\begin{lemma}
    \label{lem:minig}
    Let $G$ be a planar graph and $(F, \mathcal{P})$ a cloud decomposition of
    $G$. We can output all mini graphs of $G$ in $O(n)$ time.
\end{lemma}
}
\lemminig
\newcommand{\proofminig}{
\begin{proof}
    The idea is to search for a balanced separator $S$ of $F$ that partitions
    $V(F)$ into three parts $\{A, S, B\}$.  Continue recursively for $F'=F[A
    \cup S]$ and $F''=F[B \cup S]$ (removing edges between nodes of $S$ from one
    of these instances). Stop the recursion once $F'$ has an overall
    weight~$<\log^{\delta}n$. The nodes $v \in V(F')$ are then expanded to their
    respective clouds $C \in \mathcal{P}$. The vertex induced subgraph of $G$
    over the vertices in all these clouds $C$ is then a mini graph that we
    output. In the following, let $F'$ be the input graph of the current
    recursive call of our procedure. We denote by $w(v)$ the weight of a node
    and by $w(F')$ the total weight of all nodes of $V(F')$. Assume for now that
    $V(F')$ contains no node of weight $w(v) > 1/3 w(F')$. In this case a
    balanced separator $S$ of $F'$ of weight $O(\sqrt {w(F')})$ can easily be
    found in $O(|V(F'|)$
    time~\cite{kleinMozes2021,doi:10.1137/0136016,Miller86}, and we can progress
    recursively as outlined. If we do not encounter a problem with nodes of
    too large weight at any point, the entire recursive algorithm takes
    $O(|V(F)|\log n)=O(n)$ time. 

    Analogous to the proof of Theorem~\ref{thm:separator} we need to consider
    the special case of a leaf and bridge node $v$ exceeding the weight limit of
    $\beta w(F')$ (other nodes can not exceed this limit). In this case the
    neighborhood $S$ of $v$ in $F'$ is a separator of $V(F')$ into $\{A, S, B\}$
    with $A=\{v\}$ and $B=V(G)\setminus(A \cup S)$. Denote by $\mathcal{S}$ the
    set of all clouds represented by nodes of $S$ and by $\mathcal{C}_v$ the
    set of all clouds represented by $v$. Note that $\mathcal{S}$ contains one
    cloud if $v$ is a meta-leaf node, and two clouds if $v$ is a meta-bridge
    node. We can now output mini graphs $G_m$ directly, with each $G_m$ being
    the vertex induced subgraph over $\mathcal{S}$ and some $C \in C_v$. We
    collect the vertices of each $G_m$ as follows. Start with a set $M$
    initialized to $M=\bigcup_{S \in \mathcal{S}}S$, i.e., the separator. Now
    iterate over $C \in \mathcal{C}_v$ and set $M=M \cup C$ if $|M \cup C| <
    \log^{\delta}n$.
    If the threshold is reached or the iteration is over, we have found a mini
    graph $G_m=G[M]$. In this case reset $M=\bigcup_{S \in \mathcal{S}}S$ and
    continue. This entire process takes $O(|\mathcal{S}| +
    |\mathcal{C}_v|)=O(\log n + w(v))=O(w(F'))$ time. The recursion now only has
    to continue for $F'[V(F')-v]$, which contains less than $2/3$ times the total
    weight of $F'$.
    
    In any case the recursion is stopped once the input graph has weight $<
    \log^{\delta}n$ at which point we have found a mini graph, which can be
    output by translating nodes of the input graph $F'$ to the respective
    clouds of $G$, which induce the mini graph $G_m$ that we want to output.
    Note that obtaining the vertices $M$ of a mini graph $G_m$ can easily be
    achieved in $O(|V(G_m)|)$ time by translating the nodes of $F'$ to their
    respective clouds in $O(\log^{\delta}n)$ time. For obtaining the edges
    between vertices of $M$ we need to create a graph $G_{d}$ which is a
    directed version of $G$ such that each vertex has bounded in-degree
    (Lemma~\ref{lem:boundindegree}). Using $G_{d}$ we are able to translate the
    vertices for each mini graph linear in the size of the mini graph. Note that
    the graph $G_{d}$ needs to be created only once as a global structure in
    $O(n)$ time and $O(n)$ bits by using a dynamic subgraph to store $G_{d}$.
    The overall runtime of all recursive calls is then $O(|V(F)|\log
    n)+O(n)=O((n/\log n)\log n)=O(n)$.
\end{proof}
}
\full{\proofminig}

The bound on the number of duplicates due to Lemma~\ref{lem:blelloch} can be
upheld with some care. \conf{For details refer to Appendix~\ref{sec:proofnumduplicate}.}

\newcommand{\lemnumduplicate}{
\begin{lemma}
    \label{lem:numduplicate}
    Let $G$ be a planar graph and $(F, \mathcal{P})$ a cloud decomposition of
    $G$.
The total number of duplicate vertices among the mini
graphs of the cloud decomposition is $O(n/\log n)$.
\end{lemma}
}
\lemnumduplicate
\newcommand{\proofnumduplicate}{
\begin{proof}
As mentioned by Blelloch et al.~\cite{10.5555/1875737.1875750} it suffices if
the separator that is found is a polylogarithmic approximation when encoding an
arbitrary separable graph, i.e., if the graph that is to be encoded admits to an
$O(n^c)$-separator theorem for some $c < 1$, it suffices to find a separator of
size $O(n^c \log^k n)=O(n^{c'})$ for some $k > 1$ and some $c' > c$. When using
the standard method of finding recursive separators, the size of each separator
is dependent on the input graph of each recursive call to the separator
algorithm. As an example consider an input graph $G'$ of size $\log^{\delta}n$,
which is the largest graph that can occur as an input during the recursion (any
smaller graph is a mini graph). When searching for a separator for $G'$ a
separator of size $O((\log^{\delta}n)^{c})$ is found for some $c < 1$. Consider
our space-efficient separator search that operates on a cloud decomposition $(F,
\mathcal{P})$ constructed for $G$, with a recursive separator search on $F$
instead of $G$. In this case, the largest graph is defined by the total weight
of the input graph $F'$ of the recursion. If the total weight is
$\log^{\delta}n$, then a separator $S$ of $F'$ of size
$O((\log^{\delta-1}n)^{c})$ with total weight $O((\log^{\delta-1}n)^{c}\log n)$
is found in $F'$. The key observation is that the $\log n$ factor at the end is
independent of the size of the input graph $F'$. As outlined, the vertices that
are contained in a separator are exactly those that are duplicate vertices in
the mini graphs of the encoding. By using our scheme to output the mini graphs
increases the number of duplicate vertices by such a large factor that it
invalidates Lemma~\ref{lem:blelloch}. 
We therefore adjust the constant $\delta$ specific to the graph class---see
below.  

Intuitively, the
encoding due to Blelloch and Farzan can encode arbitrary separable graphs. For
these separable graphs, the $\delta$ values used can become arbitrarily large to
adhere to Lemma~\ref{lem:blelloch}. Therefore, we are able to increase it. In
Lemma~\ref{lem:blelloch} Blelloch and Farzan mention a constant $\delta$. In
their publication they set $\delta=1/(1-c)$ when encoding a graph $G$ that
admits to an $O(n^c)$-separator theorem. For planar graphs $c=1/2$ and therefore
$\delta=4$. When using a recursive separator algorithm until mini graphs of size
$<\log^{\delta} n=\log^4 n$ are found, the number of total mini graphs is
$O(n/\log^4n)$, which is equal to the number of leaf nodes of the recursion tree.
This tree has depth $O(\log n)$ and as it is binary, $O(n/\log^4n)$ total
nodes. The number of duplicate vertices can then bounded by $O(n/\log^4n \cdot
(\log ^4n)^{1/2})=O(n/\log^2n)$ as per Blelloch and Farzan. When using our
space-efficient approach we have to increase the $\delta$ value to adhere to
this bound, as the number of duplicates is now bounded by $O(n/\log^4n\cdot(\log
^4n)^{1/2} \log n)$ due to the constant $O(\log n)$ factor that is independent
of the smaller input graphs of the recursive separator search and only dependent
on the size of the first input graph $G$.

This would mean a total number of duplicate vertices among mini
graphs of $O(n/\log n)$. To mitigate this, we replace $\delta$ with $\delta+2$
to bound the
number of duplicates by
$O(n/\log^6n \cdot(\log^6 n)^{1/2}\log n) = O(n/\log^2 n)$. Note that arbitrary
large $\delta$ values can be chosen for the encoding as separable
graphs can be encoded with arbitrarily small $c$
values~\cite{10.5555/1875737.1875750}.
\end{proof}
}
\full{\proofnumduplicate}

The full encoding in $O(n)$ time can be done by using the algorithm of
Lemma~\ref{lem:minig} to output all mini graphs. For every mini graph $G_m$ we
again construct a cloud decomposition in linear time and use the algorithm of
Lemma~\ref{lem:minig} to find the micro graphs of $G_m$. All micro graphs are
encoded using the table lookup scheme outlined
in~\cite{10.5555/1875737.1875750}. All other structures needed for the encoding
can be constructed in $O(n)$ time and $O(n)$ bits easily, as they are simple
indexable dictionaries over vertices of the micro graphs and mini graphs or
small lookup tables which can be initialized in time linear in their size, which
is $O(n)$. 
From this description 
the next theorem follows.
\begin{theorem}
    \label{thm:encodeon}
    A planar graph $G$ with entropy $\mathcal{H}(n)$ can be succinctly encoded
    in $\mathcal{H}(n)+o(n)$ bits such that adjacency, neighborhood, and degree
    queries are supported in constant time. The encoding can be constructed in
    $O(n)$ time using $O(n)$ bits.
\end{theorem}

\section{Applications: Planar Tree Decompositions}
\label{sec:applications2}

We next present a simple modification of the recursive
separator search of Lemma~\ref{lem:minig} using standard
techniques to output a tree decomposition of $G$. 
Let $(F, \mathcal{P})$ be a cloud decomposition of $G$.
Each balanced separator $S$
of $F$ induces a bag in a tree decomposition of $F$ via the following recursive
relation. Let $F'$ be the input graph used in the recursive calls, initially
$F'=F$. Additionally, maintain a set $X$ containing vertices contained in
separators found in previous recursive calls, initially $X=\emptyset$. When a
separator $S$ is found for $F$ that partitions $V(F)$ into three sets $\{A, S,
B\}$ output the next bag of the tree decomposition of $F$ as $S \cup X$.
Continue the recursion for the input $F'[S \cup A]$ and $X=(X\cup S)\cap A$ and
the input $F'[S \cup B]$ and $X=(X\cup S)\cap B$. Note that this tree
decomposition of $F$ has width $O(\sqrt{|V(F)|})$. This tree decomposition can
easily be expanded to a tree decomposition of $G$ by expanding all vertices of a
bag $B$ to their respective clouds. The width of this expanded tree
decomposition is $O(\sqrt{|V(F)|}\log n)=O(n^{1/2+\epsilon})$ for any $\epsilon
> 0$. Each bag can be output in $O(n^{1/2+\epsilon})$ time and there are $O(n)$
bags. 
This description allows the
next corollary. 
\begin{corollary}
    \label{cor:treedecomp}
    Let $G$ be a planar graph. We can compute and output the bags of a tree
    decomposition with width $O(n^{1/2+\epsilon})$ for any $\epsilon > 0$ in
    time linear in the size of the tree decomposition $O(n^{3/2+\epsilon})$
    using $O(n)$ bits.
\end{corollary}

\section{Generalizing to {$H$-Minor-Free Graphs}}
\label{sec:general}
To generalize the previous results to {$H$-minor-free graphs
from planar graphs} we must generalize cloud partitions and their induced
minors. Recall that a cloud partition $\mathcal{P}$ for a planar graph $G$
contains $O(n/\log n)$ critical clouds. Critical clouds are exactly those of
size $< c\log n$ and with $\geq 3$ adjacent big clouds.  We define a
$\phi$-critical cloud as a small cloud adjacent to $\geq \phi$ big clouds. 
The idea of the proof is the same as for the proof of Lemma~\ref{lem:critical},
but we now have a bound $\phi$ that depends on the graph class. The existence of
the bound $\phi$ is due to the face that a separable graph $G=(V, E)$ has bound
density $d$, i.e., $|E|\leq d |V|$ for some fixed constant $d$. In particular we
show that setting $\phi$ as $d+1$ gives us the desired properties outlined in
the following. \conf{The full proof can be found at
Appendix~\ref{sec:proofphibound}.}

\newcommand{\lemphibound}{
\begin{lemma}
    \label{lem:phibound}
    Let $\mathcal{G}$ be a $H$-minor-free family of graphs for some fixed graph
    $H$ and let $d$ be the maximal density of a graph in $\mathcal{G}$. There
    exists a cloud partition $\mathcal{P}$ for each graph $G \in \mathcal{G}$
    such that $\mathcal{P}$ contains $O(n/\log n)$ $\phi$-critical clouds for
    $\phi=d+1$.
\end{lemma}
}
\lemphibound
\newcommand{\proofphibound}{
\begin{proof}
Let $\mathcal{P}$ be a cloud partition of a graph $G \in \mathcal{G}$ and
let $F=(V', E')$ be the graph constructed by contracting the vertices of each
cloud $C \in \mathcal{P}$ to a single vertex $v \in V'$. Two vertices $u, v \in
V'$ are adjacent in $F$ exactly if $C_v$ and $C_u$ are adjacent. We refer to the
vertices of $F$ introduced for $\phi$-critical clouds as $\phi$-critical
vertices and vertices introduced for big clouds as big vertices. Recall that a
$\phi$-critical cloud is a cloud of size $< \lceil c\log n \rceil$ for some
chosen constant $c$ with $\geq \phi$ adjacent big clouds. Remove all vertices
that are not big vertices or $\phi$-critical vertices and all edges incident to
two big vertices from $F$. As $\mathcal{G}$ is minor-closed and $F$ is a minor
of $G$, $F$ must be part of $\mathcal{G}$ as well. It follows that $|E'|\leq
d|V'|$ with $d$ being the maximal density of any graph part of $\mathcal{G}$.
Denote by $h$ the number of $\phi$-critical vertices and $k$ the number of big
vertices in $F$. As the degree of each $\phi$-critical vertex is $\geq \phi$ it
holds that $|E'|\geq \phi h$. Using this we can bound the number of
$\phi$-critical vertices in $F$ via $\phi h \leq d|V'|=dh+dk$ and thus $h \leq
dk/(\phi - d)$. For $\phi = d + 1$ it then follows $h \leq dk$. As there are
$O(n/\log n)$ big clouds in $\mathcal{P}$ it follows that there are $O(n/\log
n)$ $\phi$-critical clouds in $\mathcal{P}$.
\end{proof}
}
\full{\proofphibound}

It remains to show how small clouds adjacent to $<\phi$ big clouds are to be
handled. For planar graphs such clouds are exactly the bridge and leaf clouds.
{To generalize from planar graphs to $H$-minor-free graphs} we must
in turn generalize bridge clouds similarly 
to the generalization from critical to $\phi$-critical clouds. We call such
generalized bridge clouds~{\it$\phi$-bridge clouds}, which we define as small
clouds adjacent to $<\phi$, but $>2$ big clouds. Analogous to regular bridge
clouds, there can be $O(n)$ such clouds in a cloud partition for an
$H$-minor-free graph. We refer to cloud partitions with the additional labeling
of clouds as $\phi$-critical and $\phi$-bridge as a \textit{generalized cloud
partition $\mathcal{P}$}. 

To construct the structure-maintaining minor $F$
induced by a generalized cloud partition $\mathcal{P}$ we can use the same
strategy for bridge clouds and leaf clouds. For $\phi$-critical clouds the
strategy also remains the same as for critical clouds in regular cloud
partitions. For $\phi$-bridge clouds we introduce a generalized version of the
meta-bridge node to $F$, which we call $\phi$-meta-bridge node. Note that each
$\phi$-meta-bridge node has degree $< \phi$ in $F$. For adding all $\phi$-meta
bridge nodes we iterate over $i \in (3, \ldots, \phi-1)$. For each $i$ we add
all $\phi$-meta bridges with degree $i$ to $V(F)$. Adding all degree-$i$
$\phi$-meta bridge nodes $v$ takes $O(i n)$ time as for each neighbor of $v$ the
respective clouds must be explored. Thus, an overall time of $O(\phi^2 n)$ is
needed during the construction. {As described above
Lemma~\ref{lem:constructf}, we store spanning trees for meta-bridge nodes and
meta-leaf nodes.} We store and construct the exact same spanning trees for the
$\phi$-meta-bridge nodes. These spanning trees are stored in $c$ different
dynamic subgraphs, with {$c$ being some constant}, as described in
Lemma~\ref{lem:boundindegree}. Storing these dynamic
\mbox{subgraphs takes $O(c n)=O(\phi n)$ bits of space.}

\begin{lemma}
    {Let $G$ be a $H$-minor-free graph for some fixed graph $H$}
    with a generalized cloud partition $\mathcal{P}$. We can construct the minor
    induced by $\mathcal{P}$ in $O(\phi^2 n)$ time and $O(\phi n)$ bits with
    $\phi$ being a graph class dependent constant.
    
\end{lemma}

Analogous for planar graphs, we call the combination of a generalized cloud
partition $\mathcal{P}$ and a minor induced by $\mathcal{P}$ a
\textit{$\phi$-cloud decomposition}. This allows us to generalize
Theorem~\ref{thm:separator}, Theorem~\ref{thm:encoding} and
Corollary~\ref{cor:treedecomp} to {$H$-minor-free} graphs using such
generalized cloud partitions. For generalizing the proofs of these theorems and
the corollary we only must handle the new case of $\phi$-meta-bridge nodes
exceeding the weight threshold during the separator search, as all other cases
work analogous. Recall that we must handle the case when a node $v \in V(F)$ has
too large of a weight, and thus no balanced separator can be found. A new
additional case now arises when a $\phi$-meta-bridge node $v$ exceeds the weight
threshold which is handled exactly the same as meta-bridge and meta-leaf nodes.
In detail, the neighborhood $S$ of $v$ in $F$ is a separator that separates $v$
from $V(F)\setminus S$. Expanding $S$ to a separator $S'$ of $G$ then separates all
vertices in clouds represented by $v$ from the rest of the graph. The balanced
separator $S'$ then contains $O(\phi \log n)$ vertices. {For the
following theorems we make use of the linear time $O(n^{2/3})$-separator theorem
for $H$-minor-free graphs~\cite{5670816}.

\begin{theorem}
    {Let $G$ be a $H$-minor-free graph for some fixed graph $H$},
    let $\phi$ be a graph class dependent constant and $(F, \mathcal{P})$ a
    generalized cloud partition of $G$. We can compute a balanced separator $S$
    of $G$ with size $O(n^{2/3+\epsilon})$ for any $\epsilon > 0$ in $O(n/\log
    n)$ time using $O(n)$ bits.
\end{theorem}

\begin{theorem}
    A $H$-minor-free graph for some fixed graph $H$ with entropy
    $\mathcal{H}(n)$ can be succinctly encoded in $\mathcal{H}(n)+o(n)$ bits
    such that adjacency, neighborhood, and degree queries are supported in
    constant time. The construction of the encoding of takes $O(\phi^2
    n)$ time and uses $O(\phi n)$ bits with $\phi$ a graph
    class dependent constant.
\end{theorem}

\begin{corollary}
    {Let $G$ be a $H$-minor-free graph for some fixed graph $H$}, $\phi$ a graph
    class dependent constant and $(F, \mathcal{P})$ a generalized cloud
    partition of $G$. We can compute a tree decomposition of $G$ with width
    $O(n^{2/3+\epsilon})$ for any $\epsilon > 0$ in $O(n^{5/3+\epsilon})$ time using
    $O(n)$ bits.
\end{corollary}}
\full{

\section{Experimental Analysis of Cloud Decompositions}\label{sec:pract}

\newcommand{\figureone}{
\begin{figure}[ht!]
    \begin{subfigure} {0.5\textwidth}
        \includegraphics[width=\textwidth]{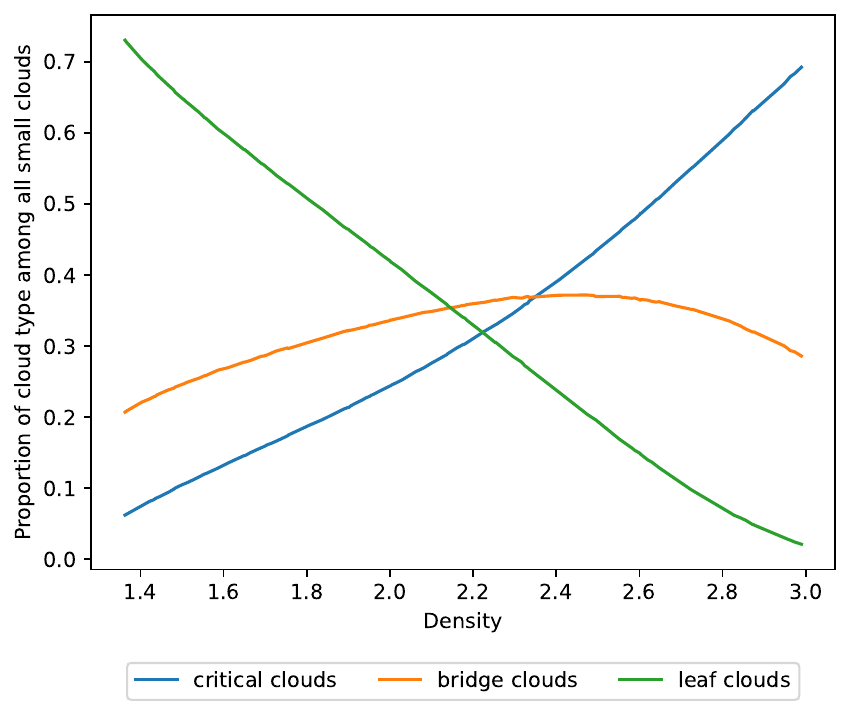} 
        \caption{Planar graphs with about 1M vertices.} \label{fig:planarb}
    \end{subfigure}
    \begin{subfigure} {0.5\textwidth}
        \includegraphics[width=\textwidth]{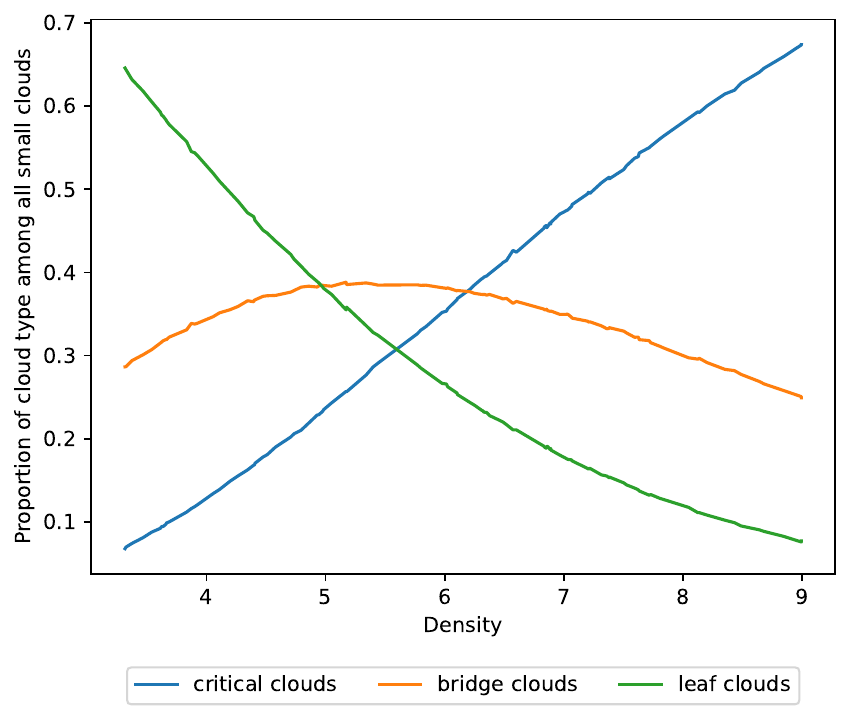} 
        \caption{Hyperbolic graphs.} \label{fig:hyperbolicb}
    \end{subfigure}
    \caption{Distribution of critical, bridge and leaf clouds among small clouds in tested graph types.}
  \end{figure}
}

\newcommand{\figuretwo}{\begin{figure}[ht!]
    \begin{subfigure} {0.5\textwidth}
        \includegraphics[width=\textwidth]{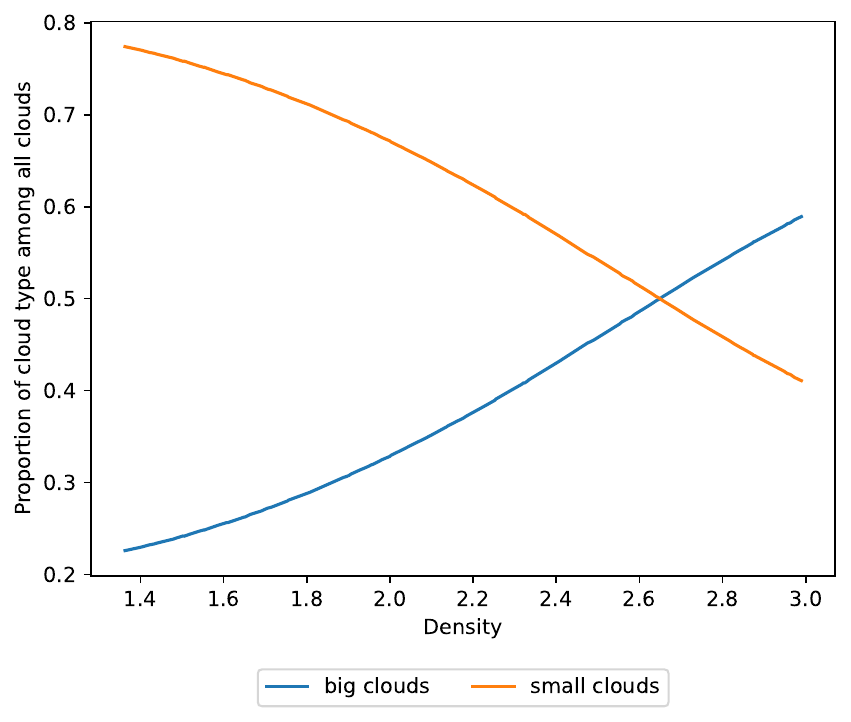}
        \caption{Planar graphs with about 10M vertices,} \label{fig:planara}
    \end{subfigure}
    \begin{subfigure} {0.5\textwidth}
        \includegraphics[width=\textwidth]{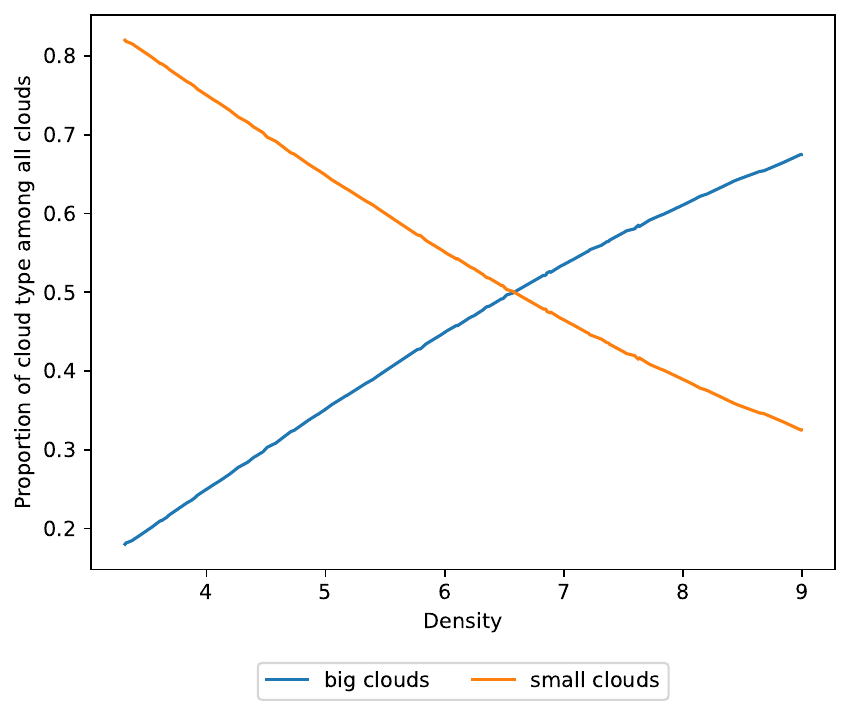} 
        \caption{Hyperbolic graphs with about 10M vertices.} \label{fig:hyperbolica}
    \end{subfigure}
    \vspace{1cm}

    \begin{subfigure} {0.5\textwidth}
        \includegraphics[width=\textwidth]{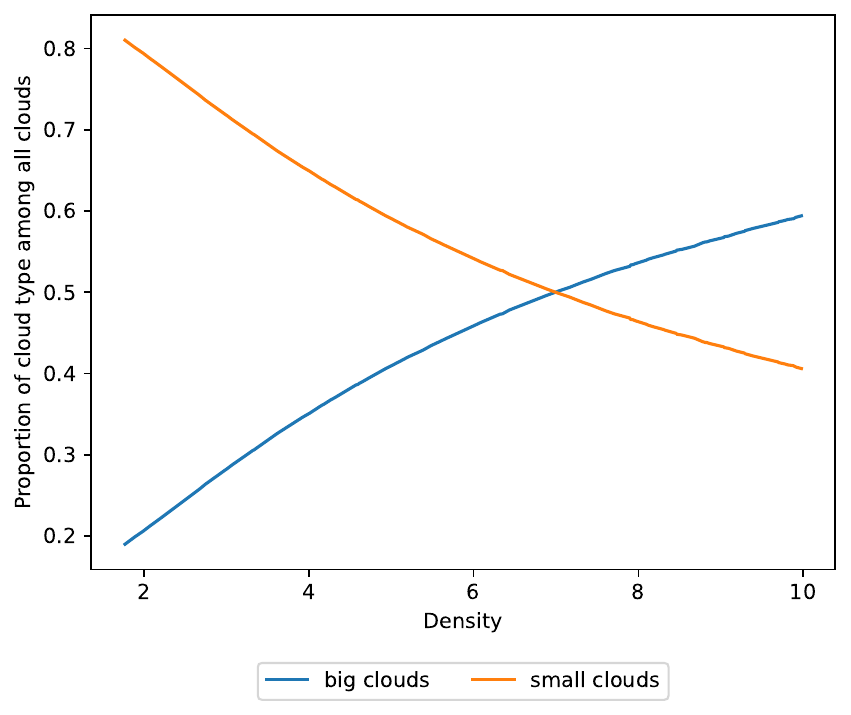} 
        \caption{G(n, p) graphs with about 10M vertices.} \label{fig:gnpa}
    \end{subfigure}
    \begin{subfigure} {0.5\textwidth}
        \includegraphics[width=\textwidth]{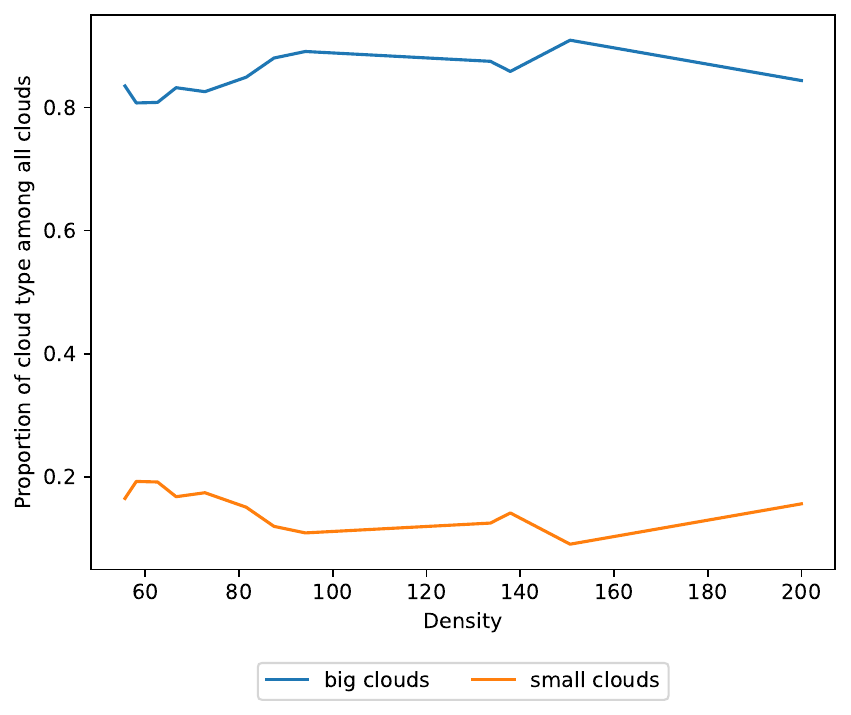} 
        \caption{Biological datasets.} \label{fig:biola}
    \end{subfigure}
    \vspace{1cm}

    \begin{subfigure} {0.5\textwidth}
        \includegraphics[width=\textwidth]{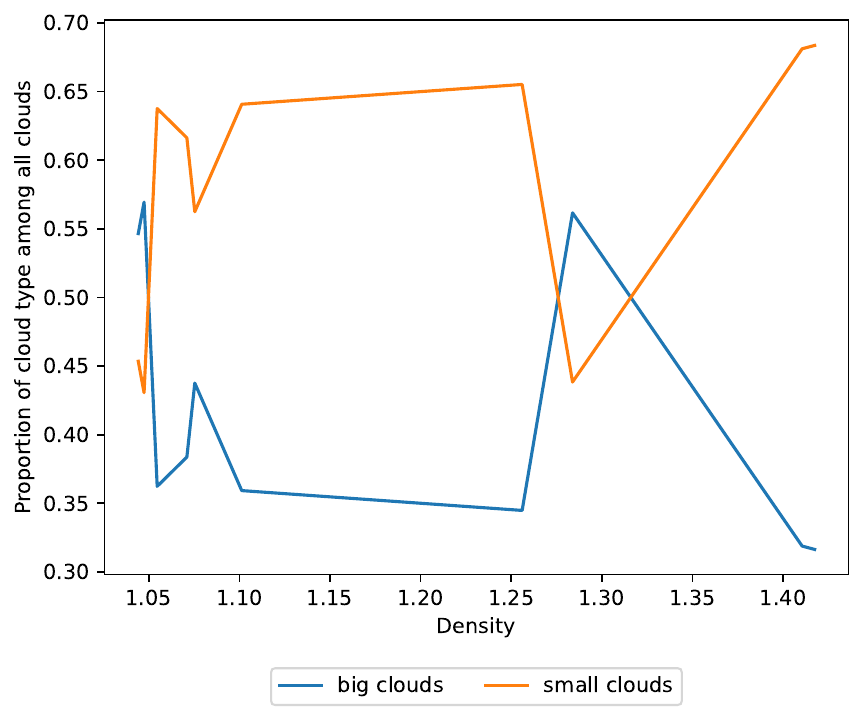} 
        \caption{Road networks.} \label{fig:roada}
    \end{subfigure}
    \begin{subfigure} {0.5\textwidth}
        \includegraphics[width=\textwidth]{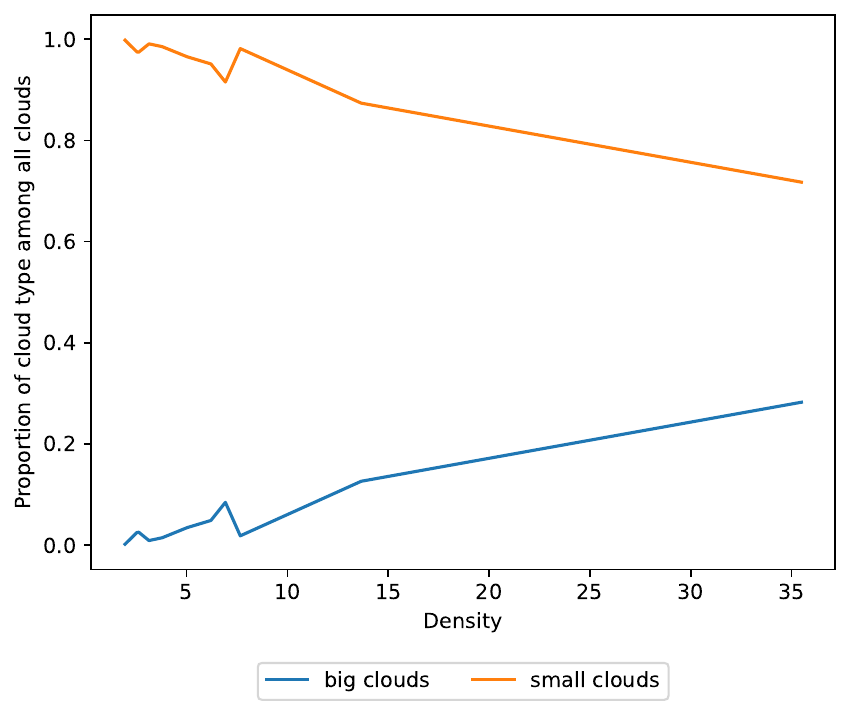} 
        \caption{Social-media graphs.} \label{fig:soca}
    \end{subfigure}
    \caption{Distribution of big and small clouds in tested graph types.}
\end{figure}
}

\newcommand{\tableone}{
    \begin{tabular}{|l|l|l||l|l|l|}
      \hline
        Name             & Big    & Small  & Critical & Bridge & Leaf                         \\ \hline
        soc-livejournal        & 8.5\%  & 91.5\% & 28.3\%   & 18.2\% & 53.6\%                       \\ \hline
        soc-delicious          & 2.5\%  & 97.5\% & 11.2\%   & 14.1\% & 74.7\%                       \\ \hline
        soc-wiki-Talk-dir      & 0.2\%  & 99.8\% & 9.1\%    & 14.4\% & 76.5\%                       \\ \hline
        soc-orkut              & 28.3\% & 71.7\% & 72.7\%   & 9.9\%  & 17.4\%                       \\ \hline
        soc-flickr             & 4.9\%  & 95.1\% & 12.8\%   & 9.5\%  & 77.8\%                       \\ \hline
        soc-pokec              & 12.6\% & 87.4\% & 48.7\%   & 17.5\% & 33.8\%                       \\ \hline
        soc-digg               & 1.8\%  & 98.2\% & 25.6\%   & 16.5\% & 57.8\%                       \\ \hline
        soc-youtube-snap       & 2.5\%  & 97.5\% & 15.2\%   & 14.2\% & 70.6\%                       \\ \hline
        soc-FourSquare         & 3.5\%  & 96.5\% & 14.0\%   & 17.1\% & 68.9\%                       \\ \hline
        soc-flixster           & 0.9\%  & 99.1\% & 20.7\%   & 15.5\% & 63.7\%                       \\ \hline
        soc-lastfm             & 1.5\%  & 98.5\% & 26.0\%   & 18.3\% & 55.7\%                      \\ \hline
        \hline \hline
        *\_M87117093 &  87.5\% & 12.5\% & 62.1\%   & 13.8\% & 24.1\%                        \\ \hline
        *\_M87128519 &  84.4\% & 15.6\% & 62.7\%   & 14.3\% & 23.0\%                        \\ \hline
        *\_M87127667 &  88.0\% & 12.0\% & 67.4\%   & 11.2\% & 21.4\%                        \\ \hline
        *\_M87102575 &  89.1\% & 10.9\% & 68.8\%   & 11.6\% & 19.6\%                        \\ \hline
        *\_M87126525 &  90.9\% & 9.1\%  & 68.7\%   & 12.1\% & 19.2\%                        \\ \hline
        *\_M87104300 &  84.9\% & 15.1\% & 63.7\%   & 13.9\% & 22.4\%                        \\ \hline
        *\_M87117515 &  83.5\% & 16.5\% & 59.1\%   & 15.2\% & 25.7\%                        \\ \hline
        *\_M87113679 &  82.6\% & 17.4\% & 61.2\%   & 14.0\% & 24.8\%                        \\ \hline
        *\_M87115834 &  83.2\% & 16.8\% & 60.8\%   & 14.3\% & 24.9\%                        \\ \hline
        *\_M87125989 &  80.8\% & 19.2\% & 56.8\%   & 15.8\% & 27.5\%                        \\ \hline
        *\_M87128519 &  84.4\% & 15.6\% & 62.7\%   & 14.3\% & 23.0\%                        \\ \hline
        *\_M87123456 &  80.7\% & 19.3\% & 59.7\%   & 15.3\% & 25.1\%                        \\ \hline
        *\_M87105849 &  85.8\% & 14.2\% & 66.1\%   & 13.4\% & 20.5\%                        \\ \hline
        \hline \hline
        road-netherlands-osm   & 35.9\% & 64.1\% & 5.8\%    & 42.8\% & 51.4\%                       \\ \hline
        road-usroads           & 56.2\% & 43.8\% & 10.1\%   & 60.1\% & 29.8\%                       \\ \hline
        road-great-britain-osm & 36.2\% & 63.8\% & 2.6\%    & 29.6\% & 67.8\%                       \\ \hline
        road-luxembourg-osm    & 54.7\% & 45.3\% & 3.2\%    & 46.1\% & 50.7\%                       \\ \hline
        road-belgium-osm       & 43.8\% & 56.2\% & 5.5\%    & 50.7\% & 43.8\%                       \\ \hline
        road-italy-osmes       & 56.9\% & 43.1\% & 4.5\%    & 49.6\% & 46.0\%                       \\ \hline
        road-usroads-48        & 56.2\% & 43.8\% & 10.1\%   & 60.1\% & 29.8\%                       \\ \hline
        road-germany-osm       & 38.4\% & 61.6\% & 3.7\%    & 38.2\% & 58.0\%                       \\ \hline
        road-euroroad          & 34.5\% & 65.5\% & 8.3\%    & 37.6\% & 54.1\%                       \\ \hline
        road-roadNet-PA        & 31.6\% & 68.4\% & 10.2\%   & 23.7\% & 66.1\%                       \\ \hline
        road-roadNet-CA        & 31.9\% & 68.1\% & 10.1\%   & 24.7\% & 65.2\%                       \\ \hline
      \end{tabular}
}

In this section we present experimental results on the distribution of cloud
types found via the cloud partition algorithm of
Section~\ref{sec:graphcoursening}. The source code of our implementation is
available on GitHub~\cite{ThmmniiiStaR2024}. Our implementation has been tested
on randomly generated planar graphs, geometric graphs, and $G(n, p)$ graphs, as
well as real-world road networks, biological datasets, and social network
graphs. For all graphs used in our experimental analysis we considered only the
largest connected component, resulting in minor fluctuations in the number of
vertices.

To generate random planar graphs, we first constructed random maximum planar
graphs using a tool from Fuentes and Navarro~\cite{ExperimentalDatasetsb} and
then removed a random number of edges between 0 and $2n$, to obtain planar
graphs having between $n$ and $3n-6$ edges. The hyperbolic graphs and the $G(n,
p)$ graphs were generated using \texttt{networkit}~\cite{networkit}. For each
graph type mentioned above we generated $100$ graphs each with $10^6$,
$5\cdot10^6$, and $10^7$ vertices, respectively. The $G(n, p)$ graphs were
generated using $1/n \leq p \leq 20/n$. From the network repository~\cite{nr},
we obtained a dataset of 11 planar road networks, each with between $114599$ and
$2216688$ vertices. For non-planar graphs, we tested 13 biological datasets,
with between $827766$ and $969582$ vertices~\cite{bigbrain,nr} and 11 social
media graphs with between $513969$ and $4033137$ vertices~\cite{nr}.

We begin by discussing our experimental results obtained for the random planar
graphs. Since critical clouds are treated exactly like big clouds when
coarsening the graph as described in Section~\ref{sec:graphcoursening}, they do
not require special attention. Therefore, instead of focussing on only small
clouds when discussing planar graphs, we give special consideration to the exact
cloud type, i.e., critical, leaf or bridge. Looking at the distribution of small
clouds in random planar graphs, i.e., critical-, bridge- and leaf
clouds, we see that among the small clouds, the fraction of critical clouds
exceeds the fraction of leaf clouds in graphs with a density of at least $2.25$,
and the fraction of bridge clouds in graphs with a density of at least $2.35$.
Figure~\ref{fig:planarb} shows the distribution among small clouds
and Figure~\ref{fig:planara} shows the distribution big
and small clouds. However, even when we encounter
many leaf and bridge clouds, the total number of clouds in the random planar
graphs is still only a small constant factor of the theoretical lower bound of
$\lceil n/\log n \rceil$. On average the total number of clouds is $2.3$ times
of the theoretical minimum of $\lceil n / \log n \rceil$ clouds, at most the\mbox{
total number of clouds is about $3$ times the theoretical lower bound.}

\arxiv{\figureone}
\elsevier{\figureone}

Looking at the results obtained for hyperbolic graphs, we can see that the
distribution of cloud types looks similar to the distribution in random planar
graphs. Analogous to planar graphs, Figure~\ref{fig:hyperbolicb} shows the
distribution among small clouds and Figure~\ref{fig:hyperbolica} shows the
distribution big and small clouds.  In hyperbolic graphs with a density of $3$
the amount of big clouds lies at around 20\%. Around a density of 6.5 the amount
of big clouds gets larger than the amount of small clouds. At a density of 9 the
amount of big clouds lies at around 70\%. The total amount of clouds is on
average about 2 times the lower bound and at worst 3 times the lower bound. 

A similar distribution can be seen for the $G(n, p)$ graphs
(Figure~\ref{fig:gnpa}). At a density of 4 there are about 30\% big clouds, at
about a density of 7 there are as many big clouds as small clouds, and at a
density of 10 there are about 60\% big clouds. The total number of clouds in
this graph type lies on average at about 2 times the lower bound, in the worst
case about 4 times the lower bound.
\elsevier{\figuretwo}
\elsevier{\FloatBarrier}
The biological datasets consist of about 90\% big clouds
(Figure~\ref{fig:biola}). The total number of clouds lies on average 13\% and at
most 18.1\% above the lower bound. The number of small clouds in the road
networks (Figure~\ref{fig:roada}) varies greatly. We can see that 8 of the 11
test graphs have more small clouds than big clouds. Despite the small number of
big clouds in some cases, the average total number of clouds is about 2 to 2.5
times the lower bound.

In contrast to the previously mentioned graph classes, the social media 
graphs behave quite
differently, as we encounter only few big clouds (Figure~\ref{fig:soca}). Comparing 
these social media graphs
with previously presented graphs of similar density, we can observe more than
45\% big clouds among those other graphs, while most social media graphs of this
density have less than 8\% big clouds, excluding two outliers with 13\% and 28\%
big clouds. Among the small clouds, leaf clouds are the most common, with a
share of 54\% - 78\% as can be seen in Table~\ref{tbl:graphs}. The total number
of clouds is far from the theoretical lower bound of $\lceil n/\log n \rceil$,
with between 6 and 17 times this lower
bound---a factor too large to consider a constant.
We explain this behavior by the high variance of vertex degrees in these graphs,
coupled with an extreme distribution of the edges. Most edges are incident to a
very small set of vertices, resulting in few vertices with very high degrees,
leaving the vast majority of vertices with low degrees. On average, 43\% of the
vertices in these graphs have a degree of 1. The graph in which this
distribution is most extreme is \textit{soc-FourSquare}, which has a median
degree of $1$ and a maximum degree of $106218$. 51\% of all vertices have degree 
1, 16\% of all vertices have degree 2.
Only about the top 0.1\% of nodes have a degree above 100. We presume that the
large number of small clouds is due to this extreme distribution of the vertex
degrees, with most vertices contributing very few edges.

To confirm this hypothesis, we randomly added vertices of degree $1$ to random
maximal planar graphs. When we add $n$ such vertices, we get planar graphs with
a density of 2. When computing a cloud decomposition on these graphs, we get
roughly 87\% bridge and leaf clouds. Planar graphs with density $2$, but no
extra degree-1 vertices, have only 48\% bridge and leaf clouds. When we add $n$
degree-2 vertices (by subdividing randomly selected edges) in random maximal
planar graphs, we get graphs with a density of $2.5$. These have about 77\%
bridge and leaf clouds while other planar graphs with a similar density have
only about 31\% bridge and leaf clouds. 
Even when adding only
$\lfloor\log n\rfloor$ vertices with degree $1$, the share of bridge and leaf
clouds lies at around 33\% with a density of 2.9. Other planar graphs with this
density have only about 16\% bridge and leaf clouds. When adding $\log n$
degree-2 vertices we get graphs with a density of 3 and roughly 12\% bridge and
leaf clouds. This is comparable to the number of bridge and leaf clouds found in
other planar graphs of similar density.

\elsevier{
    \begin{table}[hb!]
    \centering
    \scalebox{0.75}{\tableone}
    \vspace{4mm}
    \caption{Analyzed social networks, biological datasets and road networks. The share of
    big and small clouds is shown in relation to all clouds, the share of critical,
    bridge and leaf clouds in relation to all small clouds.}
    \label{tbl:graphs}
\end{table}
}
\elsevier{\FloatBarrier}
In summary, small clouds are quite rare in most graph classes, especially in
graphs with a high density and graph types where the degree distribution is
fairly well-balanced. But even for graphs with a low density, such as the road
networks and some randomly generated planar graphs and road networks, the
highest number of clouds among our test graph types is 3 times the lower bound
of $\lceil n / \log n \rceil$. We conclude that in practice it may not be
necessary to use the complex strategies that introduce meta-leafs and
meta-bridges, as there may not be enough leaf and bridge clouds to justify it.

On the other hand, graphs with a high variance in vertex degrees perform much
worse than graphs with a low variance in vertex degrees, and for these graphs
specialized strategies may be of interest, e.g., for the social-media graphs.
One such area of research would be to design preprocessing steps that low degree
vertices in a problem-specific way. Consider the problem of finding separators.
For vertices of degree $1$, it is trivial to find separators since these are
simply the sets of neighbors, and similar strategies can be used for degree $2,
3, \ldots, k$ up to some small constant $k$. If one were able to preprocess
all such vertices with degree $\leq k$, it is feasible that social-media graphs
shrink significantly, possibly by a non-constant factor. At this point, standard
techniques may be applicable while remaining space efficient. 

\arxiv{
    \begin{table}[ht!]
    \centering
    \tableone
    \vspace{5mm}
    \caption{Analyzed social networks, biological datasets and road networks. The share of
    big and small clouds is shown in relation to all clouds, the share of critical,
    bridge and leaf clouds in relation to all small clouds.}
    \label{tbl:graphs}
\end{table}
}

\arxiv{\figuretwo}

}

\clearpage
\bibliography{main}

\begin{thebibliography}{10}

\bibitem{ThmmniiiStaR2024}
thm-mni-ii/{StaR}.
\newblock URL: \url{https://github.com/thm-mni-ii/StaR}.

\bibitem{DBLP:journals/jocg/AleardiD18}
Luca~Castelli Aleardi and Olivier Devillers.
\newblock Array-based compact data structures for triangulations: Practical
  solutions with theoretical guarantees.
\newblock {\em J. Comput. Geom.}, 9(1):247--289, 2018.
\newblock \href {https://doi.org/10.20382/jocg.v9i1a8}
  {\path{doi:10.20382/jocg.v9i1a8}}.

\bibitem{DBLP:conf/cccg/AleardiDS05}
Luca~Castelli Aleardi, Olivier Devillers, and Gilles Schaeffer.
\newblock Dynamic updates of succinct triangulations.
\newblock In {\em Proceedings of the 17th Canadian Conference on Computational
  Geometry, CCCG'05, University of Windsor, Ontario, Canada, August 10-12,
  2005}, pages 134--137, 2005.

\bibitem{10.1007/11534273_13}
Luca~Castelli Aleardi, Olivier Devillers, and Gilles Schaeffer.
\newblock Succinct representation of triangulations with a boundary.
\newblock In {\em Algorithms and Data Structures}, pages 134--145. Springer,
  2005.

\bibitem{DBLP:journals/tcs/AleardiDS08}
Luca~Castelli Aleardi, Olivier Devillers, and Gilles Schaeffer.
\newblock Succinct representations of planar maps.
\newblock {\em Theor. Comput. Sci.}, 408(2-3):174--187, 2008.
\newblock \href {https://doi.org/10.1016/j.tcs.2008.08.016}
  {\path{doi:10.1016/j.tcs.2008.08.016}}.

\bibitem{Allender00thecomplexity}
Eric Allender and Meena Mahajan.
\newblock The complexity of planarity testing.
\newblock {\em Inf. Comput.}, 189(1):117--134, 2004.
\newblock \href {https://doi.org/10.1016/j.ic.2003.09.002}
  {\path{doi:10.1016/j.ic.2003.09.002}}.

\bibitem{bigbrain}
Katrin Amunts, Claude Lepage, Louis Borgeat, Hartmut Mohlberg, Timo Dickscheid,
  Marc-{\'E}tienne Rousseau, Sebastian Bludau, Pierre-Louis Bazin, Lindsay~B.
  Lewis, Ana-Maria Oros-Peusquens, Nadim~J. Shah, Thomas Lippert, Karl Zilles,
  and Alan~C. Evans.
\newblock Bigbrain: An ultrahigh-resolution 3d human brain model.
\newblock {\em Science}, 340(6139):1472--1475, 2013.
\newblock \href {https://doi.org/doi/10.1126/science.1235381}
  {\path{doi:doi/10.1126/science.1235381}}.

\bibitem{networkit}
Eugenio Angriman, Alexander van~der Grinten, Michael Hamann, Henning
  Meyerhenke, and Manuel Penschuck.
\newblock {\em Algorithms for Large-Scale Network Analysis and the NetworKit
  Toolkit}, pages 3--20.
\newblock Springer Nature Switzerland, Cham, 2022.
\newblock \href {https://doi.org/10.1007/978-3-031-21534-6_1}
  {\path{doi:10.1007/978-3-031-21534-6_1}}.

\bibitem{asano2014}
Tetsuo Asano, Taisuke Izumi, Masashi Kiyomi, Matsuo Konagaya, Hirotaka Ono,
  Yota Otachi, Pascal Schweitzer, Jun Tarui, and Ryuhei Uehara.
\newblock Depth-first search using $o(n)$ bits.
\newblock In Hee-Kap Ahn and Chan-Su Shin, editors, {\em Algorithms and
  Computation}, pages 553--564, Cham, 2014. Springer International Publishing.
\newblock \href {https://doi.org/10.1007/978-3-319-13075-0_44}
  {\path{doi:10.1007/978-3-319-13075-0_44}}.

\bibitem{auer2012graph}
Bas~Fagginger Auer and Rob~H Bisseling.
\newblock Graph coarsening and clustering on the {GPU}.
\newblock {\em Graph Partitioning and Graph Clustering}, 588:223, 2012.

\bibitem{auer2012gpu}
Bas O~Fagginger Auer and Rob~H Bisseling.
\newblock A {GPU} algorithm for greedy graph matching.
\newblock In {\em Facing the Multicore-Challenge II}, pages 108--119. Springer,
  2012.

\bibitem{10.1007/978-3-319-42634-1_10}
Niranka Banerjee, Sankardeep Chakraborty, and Venkatesh Raman.
\newblock Improved space efficient algorithms for {BFS}, {DFS} and
  applications.
\newblock In {\em Computing and Combinatorics}, pages 119--130. Springer
  International Publishing, 2016.

\bibitem{10.5555/644108.644219}
Daniel~K. Blandford, Guy~E. Blelloch, and Ian~A. Kash.
\newblock Compact representations of separable graphs.
\newblock In {\em Proceedings of the Fourteenth Annual ACM-SIAM Symposium on
  Discrete Algorithms}, SODA '03, page 679–688. Society for Industrial and
  Applied Mathematics, 2003.

\bibitem{10.5555/1875737.1875750}
Guy~E. Blelloch and Arash Farzan.
\newblock Succinct representations of separable graphs.
\newblock In {\em Proceedings of the 21st Annual Conference on Combinatorial
  Pattern Matching}, CPM 2010, page 138–150, 2010.

\bibitem{cai2021graph}
Chen Cai, Dingkang Wang, and Yusu Wang.
\newblock Graph coarsening with neural networks, 2021.
\newblock \href {http://arxiv.org/abs/2102.01350} {\path{arXiv:2102.01350}}.

\bibitem{chevalier09}
C{\'{e}}dric Chevalier and Ilya Safro.
\newblock Comparison of coarsening schemes for multilevel graph partitioning.
\newblock In Thomas St{\"{u}}tzle, editor, {\em Learning and Intelligent
  Optimization, Third International Conference, {LION} 3, Trento, Italy,
  January 14-18, 2009. Selected Papers}, volume 5851 of {\em Lecture Notes in
  Computer Science}, pages 191--205. Springer, 2009.
\newblock \href {https://doi.org/10.1007/978-3-642-11169-3\_14}
  {\path{doi:10.1007/978-3-642-11169-3\_14}}.

\bibitem{10.5555/365411.365518}
Yi-Ting Chiang, Ching-Chi Lin, and Hsueh-I Lu.
\newblock Orderly spanning trees with applications to graph encoding and graph
  drawing.
\newblock In {\em Proceedings of the Twelfth Annual ACM-SIAM Symposium on
  Discrete Algorithms}, SODA '01, page 506–515. Society for Industrial and
  Applied Mathematics, 2001.

\bibitem{Diestel07graphtheory}
Reinhard Diestel, Alexander Schrijver, and Paul~D. Seymour.
\newblock Graph theory, 2007.

\bibitem{10.1145/2591796.2591865}
Michael Elberfeld and Ken-ichi Kawarabayashi.
\newblock Embedding and canonizing graphs of bounded genus in logspace.
\newblock In {\em Proceedings of the Forty-Sixth Annual ACM Symposium on Theory
  of Computing}, STOC '14, page 383–392, New York, NY, USA, 2014. Association
  for Computing Machinery.
\newblock \href {https://doi.org/10.1145/2591796.2591865}
  {\path{doi:10.1145/2591796.2591865}}.

\bibitem{elmasry2015}
Amr Elmasry, Torben Hagerup, and Frank Kammer.
\newblock {Space-efficient Basic Graph Algorithms}.
\newblock In Ernst~W. Mayr and Nicolas Ollinger, editors, {\em 32nd
  International Symposium on Theoretical Aspects of Computer Science (STACS
  2015)}, volume~30 of {\em Leibniz International Proceedings in Informatics
  (LIPIcs)}, pages 288--301, Dagstuhl, Germany, 2015. Schloss
  Dagstuhl--Leibniz-Zentrum fuer Informatik.
\newblock URL: \url{http://drops.dagstuhl.de/opus/volltexte/2015/4921}, \href
  {https://doi.org/10.4230/LIPIcs.STACS.2015.288}
  {\path{doi:10.4230/LIPIcs.STACS.2015.288}}.

\bibitem{pmlr-v119-fahrbach20a}
Matthew Fahrbach, Gramoz Goranci, Richard Peng, Sushant Sachdeva, and Chi Wang.
\newblock Faster graph embeddings via coarsening.
\newblock In {\em Proceedings of the 37th International Conference on Machine
  Learning}, volume 119 of {\em Proceedings of Machine Learning Research},
  pages 2953--2963. PMLR, 13--18 Jul 2020.
\newblock URL: \url{https://proceedings.mlr.press/v119/fahrbach20a.html}.

\bibitem{Floreskul_Tretyakov_Dumas_2014}
Volodymyr Floreskul, Konstantin Tretyakov, and Marlon Dumas.
\newblock Memory-efficient fast shortest path estimation in large social
  networks.
\newblock {\em Proceedings of the International AAAI Conference on Web and
  Social Media}, 8:91--100, 2014.
\newblock URL: \url{https://ojs.aaai.org/index.php/ICWSM/article/view/14532}.

\bibitem{10.1137/0216064}
Greg~N. Frederickson.
\newblock Fast algorithms for shortest paths in planar graphs, with
  applications.
\newblock {\em SIAM J. Comput.}, 16(6):1004–1022, 1987.
\newblock \href {https://doi.org/10.1137/0216064} {\path{doi:10.1137/0216064}}.

\bibitem{DBLP:journals/jcss/Frederickson87}
Greg~N. Frederickson.
\newblock Upper bounds for time-space trade-offs in sorting and selection.
\newblock {\em J. Comput. Syst. Sci.}, 34(1):19--26, 1987.
\newblock \href {https://doi.org/10.1016/0022-0000(87)90002-X}
  {\path{doi:10.1016/0022-0000(87)90002-X}}.

\bibitem{ExperimentalDatasetsb}
Jos{\'{e}} Fuentes{-}Sep{\'{u}}lveda and Gonzalo Navarro.
\newblock Experimental datasets.
\newblock URL:
  \url{http://www.inf.udec.cl/~jfuentess/datasets/code.php#code_graphs}.

\bibitem{GILBERT1984391}
John~R Gilbert, Joan~P Hutchinson, and Robert~Endre Tarjan.
\newblock A separator theorem for graphs of bounded genus.
\newblock {\em Journal of Algorithms}, 5(3):391--407, 1984.
\newblock \href {https://doi.org/10.1016/0196-6774(84)90019-1}
  {\path{doi:10.1016/0196-6774(84)90019-1}}.

\bibitem{10.1006/jcss.1995.1076}
Michael~T. Goodrich.
\newblock Planar separators and parallel polygon triangulation.
\newblock {\em J. Comput. Syst. Sci.}, 51(3):374–389, dec 1995.
\newblock \href {https://doi.org/10.1006/jcss.1995.1076}
  {\path{doi:10.1006/jcss.1995.1076}}.

\bibitem{DBLP:conf/mfcs/Hagerup19}
Torben Hagerup.
\newblock A constant-time colored choice dictionary with almost robust
  iteration.
\newblock In {\em 44th International Symposium on Mathematical Foundations of
  Computer Science, {MFCS} 2019, August 26-30, 2019, Aachen, Germany}, volume
  138 of {\em LIPIcs}, pages 64:1--64:14. Schloss Dagstuhl - Leibniz-Zentrum
  f{\"{u}}r Informatik, 2019.
\newblock \href {https://doi.org/10.4230/LIPIcs.MFCS.2019.64}
  {\path{doi:10.4230/LIPIcs.MFCS.2019.64}}.

\bibitem{hagerup2020}
Torben Hagerup.
\newblock Space-efficient dfs and applications to connectivity problems:
  Simpler, leaner, faster.
\newblock {\em Algorithmica}, 82(4):1033--1056, 2020.
\newblock \href {https://doi.org/10.1007/s00453-019-00629-x}
  {\path{doi:10.1007/s00453-019-00629-x}}.

\bibitem{HAGERUP201916}
Torben Hagerup, Frank Kammer, and Moritz Laudahn.
\newblock Space-efficient {Euler} partition and bipartite edge coloring.
\newblock {\em Theoretical Computer Science}, 754:16--34, 2019.
\newblock Algorithms and Complexity.
\newblock \href {https://doi.org/10.1016/j.tcs.2018.01.008}
  {\path{doi:10.1016/j.tcs.2018.01.008}}.

\bibitem{DBLP:journals/corr/cs-DS-0101021}
Xin He, Ming{-}Yang Kao, and Hsueh{-}I Lu.
\newblock A fast general methodology for information-theoretically optimal
  encodings of graphs.
\newblock {\em {SIAM} J. Comput.}, 30(3):838--846, 2000.
\newblock \href {https://doi.org/10.1137/S0097539799359117}
  {\path{doi:10.1137/S0097539799359117}}.

\bibitem{10.1145/2459976.2459984}
Emilie Hogan, John~R. Johnson, and Mahantesh Halappanavar.
\newblock Graph coarsening for path finding in cybersecurity graphs.
\newblock In {\em Proceedings of the Eighth Annual Cyber Security and
  Information Intelligence Research Workshop}, CSIIRW '13. Association for
  Computing Machinery, 2013.
\newblock \href {https://doi.org/10.1145/2459976.2459984}
  {\path{doi:10.1145/2459976.2459984}}.

\bibitem{10.1145/321850.321852}
John Hopcroft and Robert Tarjan.
\newblock Efficient planarity testing.
\newblock {\em J. ACM}, 21(4):549–568, oct 1974.
\newblock \href {https://doi.org/10.1145/321850.321852}
  {\path{doi:10.1145/321850.321852}}.

\bibitem{6597770}
Tatsuya Imai, Kotaro Nakagawa, Aduri Pavan, N~Variyam Vinodchandran, and Osamu
  Watanabe.
\newblock An {$O(n^{1/2+\epsilon})$}-space and polynomial-time algorithm for
  directed planar reachability.
\newblock In {\em 2013 IEEE Conference on Computational Complexity}, pages
  277--286. IEEE, 2013.
\newblock \href {https://doi.org/10.1109/CCC.2013.35}
  {\path{doi:10.1109/CCC.2013.35}}.

\bibitem{izumi_et_al:LIPIcs:2020:12474}
Taisuke Izumi and Yota Otachi.
\newblock {Sublinear-Space Lexicographic Depth-First Search for Bounded
  Treewidth Graphs and Planar Graphs}.
\newblock In {\em 47th International Colloquium on Automata, Languages, and
  Programming (ICALP 2020)}, volume 168 of {\em Leibniz International
  Proceedings in Informatics (LIPIcs)}, pages 67:1--67:17. Schloss
  Dagstuhl--Leibniz-Zentrum f{\"u}r Informatik, 2020.
\newblock \href {https://doi.org/10.4230/LIPIcs.ICALP.2020.67}
  {\path{doi:10.4230/LIPIcs.ICALP.2020.67}}.

\bibitem{DBLP:journals/corr/KammerKL16}
Frank Kammer, Dieter Kratsch, and Moritz Laudahn.
\newblock Space-efficient biconnected components and recognition of outerplanar
  graphs.
\newblock {\em Algorithmica}, 81(3):1180--1204, 2019.
\newblock \href {https://doi.org/10.1007/s00453-018-0464-z}
  {\path{doi:10.1007/s00453-018-0464-z}}.

\bibitem{DBLP:journals/corr/abs-1907-00676}
Frank Kammer, Johannes Meintrup, and Andrej Sajenko.
\newblock Space-efficient vertex separators for treewidth.
\newblock {\em CoRR}, abs/1907.00676, 2019.
\newblock \href {http://arxiv.org/abs/1907.00676} {\path{arXiv:1907.00676}}.

\bibitem{kammer_et_al}
Frank Kammer and Andrej Sajenko.
\newblock {Simple $2^f$-Color Choice Dictionaries}.
\newblock In {\em 29th International Symposium on Algorithms and Computation
  (ISAAC 2018)}, volume 123 of {\em Leibniz International Proceedings in
  Informatics (LIPIcs)}, pages 66:1--66:12, Dagstuhl, Germany, 2018. Schloss
  Dagstuhl--Leibniz-Zentrum fuer Informatik.
\newblock \href {https://doi.org/10.4230/LIPIcs.ISAAC.2018.66}
  {\path{doi:10.4230/LIPIcs.ISAAC.2018.66}}.

\bibitem{1383165}
G.~Karypis and V.~Kumar.
\newblock Analysis of multilevel graph partitioning.
\newblock In {\em Supercomputing '95:Proceedings of the 1995 ACM/IEEE
  Conference on Supercomputing}, pages 29--29, 1995.
\newblock \href {https://doi.org/10.1109/SUPERC.1995.242800}
  {\path{doi:10.1109/SUPERC.1995.242800}}.

\bibitem{Karypis98afast}
George Karypis and Vipin Kumar.
\newblock A fast and high quality multilevel scheme for partitioning irregular
  graphs.
\newblock {\em SIAM JOURNAL ON SCIENTIFIC COMPUTING}, 20(1):359--392, 1998.

\bibitem{5670816}
Ken-ichi Kawarabayashi and Bruce Reed.
\newblock A separator theorem in minor-closed classes.
\newblock In {\em 2010 IEEE 51st Annual Symposium on Foundations of Computer
  Science}, pages 153--162, 2010.
\newblock \href {https://doi.org/10.1109/FOCS.2010.22}
  {\path{doi:10.1109/FOCS.2010.22}}.

\bibitem{10.1016/0166-218X(93)E0150-W}
Kenneth Keeler and Jeffery Westbrook.
\newblock Short encodings of planar graphs and maps.
\newblock {\em Discrete Appl. Math.}, 58(3):239–252, 1995.
\newblock \href {https://doi.org/10.1016/0166-218X(93)E0150-W}
  {\path{doi:10.1016/0166-218X(93)E0150-W}}.

\bibitem{kleinMozes2021}
Philip Klein and Shay Mozes.
\newblock Separators in planar graphs, 2021.
\newblock URL:
  \url{http://planarity.org/Klein_separators_in_planar_graphs.pdf}.

\bibitem{10.1145/2488608.2488672}
Philip~N. Klein, Shay Mozes, and Christian Sommer.
\newblock Structured recursive separator decompositions for planar graphs in
  linear time.
\newblock In {\em Proceedings of the Forty-Fifth Annual ACM Symposium on Theory
  of Computing}, STOC '13, page 505–514, New York, NY, USA, 2013. Association
  for Computing Machinery.
\newblock \href {https://doi.org/10.1145/2488608.2488672}
  {\path{doi:10.1145/2488608.2488672}}.

\bibitem{21958}
T.~Leighton and S.~Rao.
\newblock An approximate max-flow min-cut theorem for uniform multicommodity
  flow problems with applications to approximation algorithms.
\newblock In {\em [Proceedings 1988] 29th Annual Symposium on Foundations of
  Computer Science}, pages 422--431, 1988.
\newblock \href {https://doi.org/10.1109/SFCS.1988.21958}
  {\path{doi:10.1109/SFCS.1988.21958}}.

\bibitem{Lipton1977GeneralizedND}
Richard~J. Lipton, Donald~J. Rose, and Robert~E. Tarjan.
\newblock Generalized nested dissection.
\newblock {\em SIAM Journal on Numerical Analysis}, 16:346--358, 1977.

\bibitem{doi:10.1137/0136016}
Richard~J. Lipton and Robert~Endre Tarjan.
\newblock A separator theorem for planar graphs.
\newblock {\em SIAM Journal on Applied Mathematics}, 36(2):177--189, 1979.
\newblock \href {https://doi.org/10.1137/0136016} {\path{doi:10.1137/0136016}}.

\bibitem{Miller86}
Gary~L. Miller.
\newblock Finding small simple cycle separators for 2-connected planar graphs.
\newblock {\em J. Comput. Syst. Sci.}, 32(3):265--279, 1986.
\newblock \href {https://doi.org/10.1016/0022-0000(86)90030-9}
  {\path{doi:10.1016/0022-0000(86)90030-9}}.

\bibitem{10.1145/256292.256294}
Gary~L. Miller, Shang-Hua Teng, William Thurston, and Stephen~A. Vavasis.
\newblock Separators for sphere-packings and nearest neighbor graphs.
\newblock {\em J. ACM}, 44(1):1–29, January 1997.
\newblock \href {https://doi.org/10.1145/256292.256294}
  {\path{doi:10.1145/256292.256294}}.

\bibitem{doi:10.1137/S0097539799364092}
J.~Ian Munro and Venkatesh Raman.
\newblock Succinct representation of balanced parentheses and static trees.
\newblock {\em SIAM Journal on Computing}, 31(3):762--776, 2001.
\newblock \href {https://doi.org/10.1137/S0097539799364092}
  {\path{doi:10.1137/S0097539799364092}}.

\bibitem{10.1007/3-540-61142-8_588}
Fran{\c{c}}ois Pellegrini and Jean Roman.
\newblock Scotch: A software package for static mapping by dual recursive
  bipartitioning of process and architecture graphs.
\newblock In {\em High-Performance Computing and Networking}, pages 493--498.
  Springer Berlin Heidelberg, 1996.

\bibitem{10.1145/1290672.1290680}
Rajeev Raman, Venkatesh Raman, and Srinivasa~Rao Satti.
\newblock Succinct indexable dictionaries with applications to encoding k-ary
  trees, prefix sums and multisets.
\newblock {\em ACM Trans. Algorithms}, 3(4):43–es, November 2007.
\newblock \href {https://doi.org/10.1145/1290672.1290680}
  {\path{doi:10.1145/1290672.1290680}}.

\bibitem{reingold}
Omer Reingold.
\newblock Undirected st-connectivity in log-space.
\newblock {\em Electronic Colloquium on Computational Complexity (ECCC)}, 01
  2004.
\newblock \href {https://doi.org/10.1145/1060590.1060647}
  {\path{doi:10.1145/1060590.1060647}}.

\bibitem{nr}
Ryan~A. Rossi and Nesreen~K. Ahmed.
\newblock The network data repository with interactive graph analytics and
  visualization.
\newblock In {\em Proceedings of the Twenty-Ninth AAAI Conference on Artificial
  Intelligence}, AAAI'15, page 4292–4293. AAAI Press, 2015.
\newblock \href {https://doi.org/10.5555/2888116.2888372}
  {\path{doi:10.5555/2888116.2888372}}.

\bibitem{10.1007/978-3-319-38851-9_20}
Peter Sanders and Christian Schulz.
\newblock Advanced multilevel node separator algorithms.
\newblock In {\em Experimental Algorithms}, pages 294--309. Springer
  International Publishing, 2016.

\bibitem{strasser_et_al:LIPIcs:2020:12947}
Ben Strasser, Dorothea Wagner, and Tim Zeitz.
\newblock {Space-Efficient, Fast and Exact Routing in Time-Dependent Road
  Networks}.
\newblock In {\em 28th Annual European Symposium on Algorithms (ESA 2020)},
  volume 173 of {\em Leibniz International Proceedings in Informatics
  (LIPIcs)}, pages 81:1--81:14. Schloss Dagstuhl--Leibniz-Zentrum f{\"u}r
  Informatik, 2020.
\newblock \href {https://doi.org/10.4230/LIPIcs.ESA.2020.81}
  {\path{doi:10.4230/LIPIcs.ESA.2020.81}}.

\end{thebibliography}
\conf{
\clearpage
\phantomsection

\addcontentsline{toc}{chapter}{Appendix}
\section*{Appendix}
\appendix
\section{Constructing a directed graph with bounded in-degree}
\label{sec:proofbound}
\setcounter{lemma}{2}
\lemboundindegree
\proofboundindegree
\section{Constructing a cp-structure}
\label{sec:proofcpstruct}
\setcounter{lemma}{8}
\lemcpstruct
\proofcpstruct
\section{Constructing a structure-maintaining minor induced by a cloud partition}
\label{sec:proofconstructf}
\setcounter{lemma}{9}
\lemconstructf
\proofconstructf
\section{Constructing a balanced separator with the use of a cloud decomposition}
\label{sec:proofseparator}
\setcounter{lemma}{10}
\lemseparator
\proofseparator
\section{Outputting all mini graphs}
\label{sec:proofminig}
\setcounter{lemma}{13}
\lemminig
\proofminig
\section{Bounding the number of duplicate vertices in mini graphs}
\label{sec:proofnumduplicate}
\setcounter{lemma}{14}
\lemnumduplicate
\proofnumduplicate

\section{Bounding the number of $\phi$-critical vertices}
\label{sec:proofphibound}
\setcounter{lemma}{17}
\lemphibound
\proofphibound
}
\end{document}